\documentclass[preprint,12pt]{elsarticle}

\usepackage{amssymb}

\usepackage{amsthm}

\usepackage{amsmath}
\DeclareMathOperator{\tr}{tr}

\DeclareMathOperator{\re}{Re}

\newcommand{\Slash}[1]{{\ooalign{\hfil/\hfil\crcr$#1$}}}
\numberwithin{equation}{section}
\newtheorem{lem}{Lemma}

\usepackage{url}

\begin{document}

\begin{frontmatter}



\title{Supersymmetry, chiral symmetry and the generalized BRS transformation in
lattice formulations of 4D $\mathcal{N}=1$ SYM}


\author{Hiroshi Suzuki}
\ead{hsuzuki@riken.jp}
\address{%
Theoretical Research Division, RIKEN Nishina Center, Wako 2-1, Saitama 351-0198,
Japan}

\begin{abstract}
In the context of the lattice regularization of the four-dimensional
$\mathcal{N}=1$ supersymmetric Yang--Mills theory (4D $\mathcal{N}=1$ SYM), we
formulate a generalized BRS transformation that treats the gauge,
supersymmetry (SUSY), translation and axial $U(1)$ ($U(1)_A$) transformations
in a unified way. A resultant Slavnov--Taylor identity or the Zinn-Justin
equation gives rise to a strong constraint on the quantum continuum limit of
symmetry breaking terms with the lattice regularization. By analyzing the
implications of the constraint on operator-mixing coefficients in the SUSY and
the $U(1)_A$ Ward--Takahashi (WT) identities, we prove to all orders of
perturbation theory in the continuum limit that, (i)~the chiral symmetric limit
implies the supersymmetric limit and, (ii)~a three-fermion operator that might
potentially give rise to an exotic breaking of the SUSY WT identity does not
emerge. In previous literature, only a naive or incomplete treatment on these
points can be found. Our results provide a solid theoretical basis for lattice
formulations of the 4D $\mathcal{N}=1$ SYM.
\end{abstract}

\end{frontmatter}


\section{Introduction}
\label{sec:1}

Ever since the feasibility of non-perturbative studies of 4D $\mathcal{N}=1$
SYM---the simplest but still quite non-trivial 4D supersymmetric gauge
theory---based on the lattice regularization was pointed
out~\cite{Curci:1986sm},\footnote{See also Ref.~\cite{Kaplan:1983sk} for an
earlier consideration.} there has been considerable research based on the
proposed scenario of the SUSY restoration. Classified by the method for
simulating the gaugino or gluino, a fermionic superpartner of the gauge boson,
Refs.~\cite{Montvay:1996pz,Montvay:1997ak,Koutsoumbas:1997de,Kirchner:1998nk,%
Kirchner:1998mp,Campos:1999du,Feo:1999hw,Feo:1999hx,Farchioni:2000mp,%
Farchioni:2000kb,Farchioni:2001yn,Farchioni:2001yr,Farchioni:2001wx,%
Peetz:2002sr,Farchioni:2004fy,Demmouche:2008ms,Demmouche:2008aq,%
Demmouche:2009ki,Demmouche:2010sf,Bergner:2011wf} use the Wilson
fermion~\cite{Wilson:1975hf} (see Ref.~\cite{Montvay:2001aj} for a very
readable review),
Refs.~\cite{Fleming:2000fa,Giedt:2008xm,Endres:2009yp,Endres:2009pu} employ the
domain-wall fermion~\cite{Kaplan:1992bt,Shamir:1993zy} and, most recently,
Ref.~\cite{Kim:2011fw} uses the overlap
fermion~\cite{Neuberger:1997fp,Neuberger:1998wv}. Some related works are
Refs.~\cite{Nishimura:1997vg,Maru:1997kh,Donini:1997hh,Taniguchi:1999fc,%
Kaplan:1999jn}.\footnote{Ref.~\cite{Mehta:2011ud} is discussing a possible
implication of the lattice BRS invariance (in the non-perturbative level) in
supersymmetric gauge theories.}

The central issue in the scenario~\cite{Curci:1986sm} is the restoration of
SUSY and the axial $U(1)$ symmetry, $U(1)_A$, in the quantum continuum limit;
the lattice regularization generally breaks these symmetries that define 4D
$\mathcal{N}=1$ SYM. The basic physical picture of~Ref.~\cite{Curci:1986sm} is
very simple (see also~Ref.~\cite{Kaplan:1983sk}): In terms of the effective
field theory, a unique relevant operator that breaks SUSY is the gluino mass
term. If one can have a $U(1)_A$ symmetry (either by the fine-tuning of the
bare gluino mass~\cite{Karsten:1980wd,Bochicchio:1985xa,Testa:1998ez} or by use
of the Ginsparg--Wilson relation~\cite{Ginsparg:1981bj}; see below), the
$U(1)_A$ would forbid the gluino mass term and one would end up with a
supersymmetric theory in the continuum limit. In this way, SUSY might emerge as
an accidental symmetry accompanied by the chiral symmetry. The situation is a
little bit more complicated because the $U(1)_A$ suffers from the axial
anomaly. Still, one could use a remaining $\mathbb{Z}_{2N_c}\subset U(1)_A$
symmetry\footnote{Throughout this paper, we assume that the gauge group
is~$SU(N_c)$.} to forbid the gluino mass term. Thus one expects that, in the
continuum limit, the restoration of SUSY and that of the chiral symmetry occur
simultaneously. 

The best characterization of a symmetry property of a quantum field theory is
given by the WT identity. Although the basic physical picture of the above
scenario is simple, to show the validity of the scenario in terms of WT
identities is not so simple.\footnote{For a lattice
formulation~\cite{Sugino:2003yb,Sugino:2004qd} of the two-dimensional
$\mathcal{N}=(2,2)$ SYM, the restoration of the SUSY WT identity without
fine-tuning can be argued perturbatively~\cite{Kadoh:2009rw} and has been
confirmed non-perturbatively~\cite{Kanamori:2008bk}.} For example, the analysis
of~Ref.~\cite{Curci:1986sm} considers a lattice ``identity'', such as (Eq.~(14)
of~Ref.~\cite{Curci:1986sm} with~$\varOmega=A_\rho(y)\Bar{\lambda}(z)$, in our
notation)
\begin{align}
   &\left\langle\partial_\mu^*S_\mu(x)A_\nu^a(y)\Bar{\psi}^b(z)\right\rangle
\notag\\
   &\stackrel{?}{=}
   \left\langle\left[M\chi(x)+ X_S(x)\right]A_\nu^a(y)\Bar{\psi}^b(z)
   \right\rangle
   -\left\langle
   \frac{1}{a^4}\frac{\partial}{\partial\Bar{\xi}(x)}\delta_\xi
   \left[A_\nu^a(y)\Bar{\psi}^b(z)\right]\right\rangle,
\label{eq:(1.1)}
\end{align}
where $S_\mu(x)$ is the supercurrent on the lattice, $X_S(x)$ is an explicit
SUSY breaking term in the lattice action, $A_\nu^a(y)$ is the gauge potential
and~$\Bar{\psi}^b(z)$ is the gluino field. $\delta_\xi$ in the last term is the
SUSY transformation on the lattice. This relation, however, does \emph{not\/}
hold even in the tree level approximation of the perturbation
theory.\footnote{I would like to thank Yusuke Taniguchi for noticing this
observation.} The point is that the inserted operator
$A_\nu^a(y)\Bar{\psi}^b(z)$ is neither gauge invariant nor covariant thus the
contribution of the gauge-fixing term (and the ghost action for higher orders)
cannot be neglected. Moreover, as found in~Ref.~\cite{Taniguchi:1999fc} through
an explicit one-loop calculation, $X_S(x)$ mixes even with gauge non-invariant
operators, a situation that was not presumed in~Ref.~\cite{Curci:1986sm}. In
these aspects, the WT identity associated with SUSY is much more complicated
than that for the chiral symmetry~\cite{Bochicchio:1985xa,Testa:1998ez} as
noted sometime ago~\cite{Farchioni:2001wx}. It is certainly unlikely that those
gauge non-invariant elements affect a physical question. Nevertheless, it is
not quite obvious whether or not these complications modify the conclusion
of~Ref.~\cite{Curci:1986sm} that both the $U(1)_A$ WT identity (with the axial
anomaly) and the SUSY WT identity are restored by a single fine-tuning of the
bare gluino mass. Since the axial $U(1)$ current and the supercurrent belong to
a single SUSY multiplet (at least in the classical
theory)~\cite{Ferrara:1974pz}, it is quite conceivable that one can relate the
$U(1)_A$ WT identity and the SUSY WT identity by considering an algebra
involving both SUSY and the $R$-symmetry, $U(1)_A$. It is not a priori clear
how to carry out this program, however, especially with the lattice
regularization that generally breaks both symmetries explicitly.

The original motivation of the present work was to obtain a transparent
understanding on the above issue. In the course of the analysis, however, we
encountered another difficulty that has not been noted in previous literature.
That is that, generally, one might have a mixing of~$X_S(x)$ with a gauge
invariant three-fermion operator. Since this three-fermion operator has the
same mass-dimension as $X_S(x)$ and same transformation properties as~$X_S(x)$
under lattice symmetries, a dimensional counting and a simple symmetry argument
alone cannot exclude the possibility of such a mixing. But then, one would have
an exotic breaking of the SUSY WT identity even in the continuum limit (and
even with the fine-tuning of the gluino mass elucidated above).
See~Eq.~\eqref{eq:(4.20)}. According to the supposed scenario of the SUSY
restoration, we should not have such an exotic breaking.

Considering recent increases in computing power that would enable a true
realization of 4D $\mathcal{N}=1$ SYM and a fundamental role that the SUSY WT
identity should play there, it must be important to clarify the above issues;
this is the objective of the present paper. We will find that the adoption of
the generalized BRS
transformation~\cite{White:1992ai,Maggiore:1995gr,Maggiore:1996gg} in the
lattice framework, that treats the gauge, SUSY, translation and $U(1)_A$
transformations in a unified way, does the required job.

This paper is organized as follows. In~Sec.~\ref{sec:2}, we summarize basic
assumptions we make on the lattice formulation. They are physically natural and
very mild; actually all the lattice formulations of 4D $\mathcal{N}=1$ SYM
considered so far are covered by our analysis. Sections.~\ref{sec:3}
and~\ref{sec:4} are mostly a review of well-known facts concerning the
operator mixing in the $U(1)_A$ WT
identity~\cite{Bochicchio:1985xa, Testa:1998ez} and the SUSY WT
identity~\cite{Curci:1986sm,Taniguchi:1999fc,Farchioni:2001wx}; however, we
tried not to make any ad hoc assumption on the possible operator-mixing
structure. Eqs.~\eqref{eq:(4.22)} and~\eqref{eq:(4.23)} summarize our main
assertions in this paper. After these rather long preparations,
in~Sec.~\ref{sec:5}, we introduce the generalized BRS transformation on lattice
variables. There, some complications inherent in the present supersymmetric
system in the Wess--Zumino (WZ) gauge are explained. After identifying
$O(a)$~breaking terms attributed to the lattice regularization, we obtain a
Wess--Zumino (WZ)-like consistency condition~\cite{Wess:1971yu} by using the
algebraic property of the generalized BRS transformation. In~Sec.~\ref{sec:6},
finally, combining the consistency condition and the general structure of the
operator mixing explored in~Secs.~\ref{sec:3} and~\ref{sec:4}, we prove the
main assertions, Eqs.~\eqref{eq:(4.22)} and~\eqref{eq:(4.23)}. Sec.~\ref{sec:7}
concludes the paper. Appendices contain some useful information required in the
main text.

\section{Basic lattice framework}
\label{sec:2}

Our arguments below are quite general and rather independent of a particular
lattice formulation we adopt.\footnote{We basically follow the notation
of~Ref.~\cite{Taniguchi:1999fc} except the points that $x$, $y$, $z$,
\dots\ denote lattice sites, vector indices $\mu$, $\nu$, \dots, run over $0$,
$1$, $2$, $3$ and the lattice spacing~$a$ is explicitly written. The gauge
potentials~$A_\mu(x)$ are defined from the link variables by
\begin{equation}
   U(x,\mu)=e^{iagA_\mu(x)},
\label{eq:(2.1)}
\end{equation}
where $g$ is the bare gauge coupling constant. All gamma matrices are hermitian
and obey $\{\gamma_\mu,\gamma_\nu\}=2\delta_{\mu\nu}$;
$\gamma_5\equiv-\gamma_0\gamma_1\gamma_2\gamma_3$
and~$\sigma_{\mu\nu}\equiv[\gamma_\mu,\gamma_\nu]/2$. The charge conjugation
matrix~$C$ satisfies, $C^{-1}\gamma_\mu C=-\gamma_\mu^T$,
$C^{-1}\sigma_{\mu\nu}C=-\sigma_{\mu\nu}^T$, $C^{-1}\gamma_5C=\gamma_5^T$
and~$C^T=-C$. $\epsilon_{\mu\nu\rho\sigma}$ denotes the totally antisymmetric
tensor and~$\epsilon_{0123}=-1$. The generators of the gauge group $SU(N_c)$,
$T^a$, are normalized as $\tr(T^aT^b)=(1/2)\delta^{ab}$. $\Phi^a(x)$ denotes
the gauge component of a generic field~$\Phi(x)$, $\Phi(x)\equiv\Phi^a(x)T^a$.
Throughout this article, the symbol~$\tr$ denotes the trace over gauge indices.
The forward and backward difference operators respectively are defined by
\begin{equation}
   \partial_\mu f(x)
   \equiv\frac{1}{a}\left[f(x+a\Hat{\mu})-f(x)\right],\qquad
   \partial_\mu^*f(x)
   \equiv\frac{1}{a}\left[f(x)-f(x-a\Hat{\mu})\right].
\label{eq:(2.2)}
\end{equation}
The symbol of the functional derivative with respect to a lattice variable
implies
\begin{equation}
   \frac{\delta}{\delta\Phi^a(x)}
   \equiv\frac{1}{a^4}\frac{\partial}{\partial\Phi^a(x)},
\label{eq:(2.3)}
\end{equation}
for~$\Phi^a(x)$, for example.} Our lattice action consists of the sum of a
gauge boson (gluon) action $S_{\text{gluon}}$, the kinetic term of the
gluino~$S_{\text{gluino}}$ and the mass term of the gluino,
$S_{\text{mass}}^{(0)}$. A minimal requirement for the lattice actions is that
they have the correct \emph{classical\/} continuum limit such that
\begin{equation}
   S_{\text{gluon}}\xrightarrow{a\to0}
   \Breve{S}_{\text{gluon}}\equiv
   \int d^4x\,\frac{1}{2}\tr\left[F_{\mu\nu}(x)F_{\mu\nu}(x)\right],
\label{eq:(2.4)}
\end{equation}
where $F_{\mu\nu}(x)$ is the field strength
$F_{\mu\nu}(x)\equiv\partial_\mu A_\nu(x)-\partial_\nu A_\mu(x)%
+ig[A_\mu(x),A_\nu(x)]$,
\begin{equation}
   S_{\text{gluino}}\equiv
   a^4\sum_x\tr\left[\Bar{\psi}(x)D\psi(x)\right]
   \xrightarrow{a\to0}
   \Breve{S}_{\text{gluino}}\equiv
   \int d^4x\,\tr\left[\Bar{\psi}(x)\Slash{D}\psi(x)\right],
\label{eq:(2.5)}
\end{equation}
where $D$ is a lattice Dirac operator and $\Slash{D}\equiv\gamma_\mu D_\mu$
and~$D_\mu\equiv\partial_\mu+ig[A_\mu(x),]$ is the covariant derivative with
respect to the adjoint representation, and
\begin{equation}
   S_{\text{mass}}^{(0)}\equiv
   a^4\sum_xM\tr\left[\Bar{\psi}(x)\psi(x)\right]
   \xrightarrow{a\to0}
   \Breve{S}_{\text{mass}}^{(0)}\equiv
   \int d^4x\,M\tr\left[\Bar{\psi}(x)\psi(x)\right],
\label{eq:(2.6)}
\end{equation}
where $M$ is the bare gluino mass. The gluino field~$\psi(x)$ on the Euclidean
lattice (and in the Euclidean continuum theory as well) is subject to the
constraint
\begin{equation}
   \Bar{\psi}(x)=\psi^T(x)(-C^{-1}),
\label{eq:(2.7)}
\end{equation}
to express the degrees of freedom of the Majorana fermion in the Minkowski
space. Consequently, the lattice Dirac operator should
satisfy~$(C^{-1}D)^T=-C^{-1}D$. We further assume that the lattice actions are
invariant under the hypercubic transformations, the parity
transformation~$\mathcal{P}$,
\begin{align}
   U_0(x_0,\Vec{x})&\xrightarrow{\mathcal{P}}U_0(x_0,-\Vec{x}),&
   U_k(x_0,\Vec{x})&\xrightarrow{\mathcal{P}}U_k^\dagger(x_0,-\Vec{x}-a\hat k),
\notag\\
   \psi(x_0,\Vec x)&\xrightarrow{\mathcal{P}}i\gamma_0\psi(x_0,-\Vec x),&
   \Bar{\psi}(x_0,\Vec x)&\xrightarrow{\mathcal{P}}
   -i\Bar{\psi}(x_0,-\Vec x)\gamma_0,
\label{eq:(2.8)}
\end{align}
and the time reversal transformation~$\mathcal{T}$,
\begin{align}
   U_0(x_0,\Vec{x})&\xrightarrow{\mathcal{T}}U_0^\dagger(-x_0-a,\Vec{x}),&
   U_k(x_0,\Vec{x})&\xrightarrow{\mathcal{T}}U_k(-x_0,\Vec{x}),
\notag\\
   \psi(x_0,\Vec{x})&\xrightarrow{\mathcal{T}}
   i\gamma_0\gamma_5\psi(-x_0,\Vec{x}),&
   \Bar{\psi}(x_0,\Vec{x})&\xrightarrow{\mathcal{T}}
   -i\Bar{\psi}(-x_0,\Vec{x})\gamma_5\gamma_0.
\label{eq:(2.9)}
\end{align}

The simplest lattice formulation which fulfills the above requirements is the
sum of the Wilson plaquette action,
\begin{equation}
   S_{\text{gluon}}=\sum_x\sum_{\mu,\nu}\left(-\frac{1}{g^2}\right)
   \re\tr\left[U_\mu(x)U_\nu(x+a\hat\mu)U_\mu^\dagger(x+a\hat\nu)U_\nu^\dagger(x)
   \right],
\label{eq:(2.10)}
\end{equation}
and the Wilson fermion action,
\begin{align}
   S_{\text{gluino}}&=a^4\sum_x
   \tr\left(\Bar{\psi}(x)\left\{
   \frac{1}{2}\sum_\mu\left[\gamma_\mu(\nabla_\mu+\nabla_\mu^*)
   -ra\nabla_\mu^*\nabla_\mu\right]\right\}\psi(x)\right),
\notag\\
   S_{\text{mass}}^{(0)}&=a^4\sum_x
   M\tr\left[\Bar{\psi}(x)\psi(x)\right],
\label{eq:(2.11)}
\end{align}
where lattice covariant differences in the adjoint representation are defined
by
\begin{align}
   &\nabla_\mu\psi(x)
   \equiv\frac{1}{a}
   \left[U_\mu(x)\psi(x+a\hat\mu)U_\mu^\dagger(x)-\psi(x)\right],
\notag\\
   &\nabla_\mu^*\psi(x)
   \equiv\frac{1}{a}
   \left[\psi(x)
   -U_\mu^\dagger(x-a\hat\mu)\psi(x-a\hat\mu)U_\mu(x-a\hat\mu)\right].
\label{eq:(2.12)}
\end{align}

In what follows, we often consider correlation functions containing elementary
fields that are not gauge invariant. To define such correlation functions, we
have to introduce the gauge-fixing term and the Faddeev--Popov (FP) ghost and
anti-ghost fields. We thus define
\begin{align}
   S_{\text{GF}+\text{FP}}^{(0)}&\equiv-s_0a^4\sum_x2\tr\left\{
   \Bar{c}(x)\left[\partial_\mu^*A_\mu(x)+\frac{\alpha}{2}B(x)\right]
   \right\},
\label{eq:(2.13)}
\end{align}
where $\Bar{c}(x)$ and $B(x)$ are the anti-ghost and auxiliary fields,
respectively and $\alpha$~is the gauge parameter. The nilpotent BRS
transformation~$s_0$ for lattice variables is defined by (see, for example,
Ref.~\cite{Luscher:1988sd}),
\begin{align}
   s_0A_\mu(x)&\equiv\left[D_\mu c\right]^L(x),&&
\notag\\
   s_0\psi(x)&\equiv-ig\{c(x),\psi(x)\},&
   s_0\Bar{\psi}(x)&=-ig\{c(x),\Bar{\psi}(x)\},
\notag\\
   s_0c(x)&\equiv-igc(x)^2,&&
\notag\\
   s_0\Bar{c}(x)&\equiv B(x),&&
\notag\\
   s_0B(x)&\equiv0,&&
\label{eq:(2.14)}
\end{align}
where
\begin{equation}
   \left[D_\mu c\right]^L(x)
   \equiv
   \frac{iag\varDelta_{A_\mu(x)}}{1-\exp\left[-iag\varDelta_{A_\mu(x)}\right]}
   \partial_\mu c(x)+ig\varDelta_{A_\mu(x)}c(x),
\label{eq:(2.15)}
\end{equation}
and $\varDelta_X$ is the adjoint action, $\varDelta_XY\equiv[X,Y]$. We set
\begin{equation}
   \varphi(x_0,\Vec{x})\xrightarrow{\mathcal{P}}\varphi(x_0,-\Vec{x}),\qquad
   \varphi(x_0,\Vec{x})\xrightarrow{\mathcal{T}}\varphi(-x_0,\Vec{x}),
\label{eq:(2.16)}
\end{equation}
for all $\varphi(x)\equiv c(x)$, $\Bar{c}(x)$ and~$B(x)$. Then one can verify
that $[D_\mu c]^L(x)$ in~Eq.~\eqref{eq:(2.15)} behaves in an identical way
as~$A_\mu(x)$ under $\mathcal{P}$ and~$\mathcal{T}$ and, consequently,
$s_0$~preserves transformation properties under $\mathcal{P}$
and~$\mathcal{T}$. This shows that (since the
combination~$\partial_\mu^*A_\mu(x)$ behaves as~Eq.~\eqref{eq:(2.16)})
$S_{\text{GF+FP}}^{(0)}$~\eqref{eq:(2.13)} is invariant under $\mathcal{P}$
and~$\mathcal{T}$. Our total lattice action is thus given by
\begin{equation}
   S^{(0)}\equiv S_{\text{gluon}}+S_{\text{gluino}}+S_{\text{mass}}^{(0)}
   +S_{\text{GF+FP}}^{(0)},
\label{eq:(2.17)}
\end{equation}
and the corresponding classical continuum limit is
\begin{equation}
   \Breve{S}^{(0)}\equiv\Breve{S}_{\text{gluon}}+\Breve{S}_{\text{gluino}}
   +\Breve{S}_{\text{mass}}^{(0)}+\Breve{S}_{\text{GF+FP}}^{(0)},
\label{eq:(2.18)}
\end{equation}
where $\Breve{S}_{\text{GF+FP}}^{(0)}$ is the classical continuum limit
of~$S_{\text{GF+FP}}^{(0)}$~\eqref{eq:(2.13)}.

\section{$U(1)_A$ WT identity on the lattice}
\label{sec:3}

\subsection{Derivation of the $U(1)_A$ WT identity}

Let us begin our discussion with a WT identity associated with the $U(1)_A$
transformation. We define a localized version of the $U(1)_A$ transformation on
lattice variables by
\begin{equation}
   \delta_\theta\psi(x)=i\theta(x)\gamma_5\psi(x),\qquad
   \delta_\theta\Bar{\psi}(x)=i\theta(x)\Bar{\psi}(x)\gamma_5,
\label{eq:(3.1)}
\end{equation}
and $\delta_\theta=0$ on other variables. Then, from a variation of the lattice
action (note that $S_{\text{GF}+\text{FP}}^{(0)}$~\eqref{eq:(2.13)} does not
contain the gluino field),
\begin{equation}
   \delta_\theta S^{(0)}\equiv
   a^4\sum_xi\theta(x)
   \left[-\partial_\mu^*j_{5\mu}(x)+2MP(x)+X_A(x)\right],
\label{eq:(3.2)}
\end{equation}
we obtain the divergence of the axial-vector current~$j_{5\mu}(x)$ whose
classical continuum limit is given by\footnote{A possible form
for~$j_{5\mu}(x)$ is
$j_{5\mu}(x)=\tr[\Bar{\psi}(x)\gamma_\mu\gamma_5U_\mu(x)\psi(x+a\hat\mu)%
U_\mu^\dagger(x)]$~\cite{Taniguchi:1999fc}.}
\begin{equation}
   j_{5\mu}(x)\xrightarrow{a\to0}
   \Breve{\jmath}_{5\mu}(x)\equiv
   \tr\left[\Bar{\psi}(x)\gamma_\mu\gamma_5\psi(x)\right],
\label{eq:(3.3)}
\end{equation}
the pseudo-scalar density $P(x)$, which arises from the variation of the gluino
mass term~$S_{\text{mass}}^{(0)}$~\eqref{eq:(2.11)},
\begin{equation}
   P(x)\equiv\tr\left[\Bar{\psi}(x)\gamma_5\psi(x)\right],
\label{eq:(3.4)}
\end{equation}
and $X_A(x)$, a $U(1)_A$ symmetry breaking term associated with the lattice
regularization (e.g., the Wilson term, for the case of~Eq.~\eqref{eq:(2.11)}).
In~Eq.~\eqref{eq:(3.2)}, the separation between $-\partial_\mu^*j_{5\mu}(x)$
and~$X_A(x)$ is not unique and there remains $O(a)$~ambiguity in the definition
of~$X_A(x)$ even with requirement~\eqref{eq:(3.3)}. We partially fix this
ambiguity by requiring that the breaking~$X_A(x)$ is a gauge-invariant local
combination of lattice fields that behaves in an identical way
as~$P(x)$~\eqref{eq:(3.4)} under lattice discrete transformations. In
particular, we assume that
\begin{equation}
   X_A(x_0,\Vec{x})\xrightarrow{\mathcal{P}}-X_A(x_0,-\Vec{x}),\qquad
   X_A(x_0,\Vec{x})\xrightarrow{\mathcal{T}}-X_A(-x_0,\Vec{x}).
\label{eq:(3.5)}
\end{equation}
Then by considering the variation of the lattice action~\eqref{eq:(3.2)} in the
functional integral, we find the identity
\begin{align}
   &\left\langle
   \partial_\mu^*j_{5\mu}(x)\mathcal{O}(y,z,\dotsc)
   \right\rangle
\notag\\
   &=\left\langle\left[2MP(x)+X_A(x)\right]
   \mathcal{O}(y,z,\dotsc)\right\rangle
   +i\left\langle\frac{1}{a^4}\frac{\partial}{\partial\theta(x)}
   \delta_\theta\mathcal{O}(y,z,\dotsc)\right\rangle,
\label{eq:(3.6)}
\end{align}
where $\mathcal{O}(y,z,\dotsc)$ is any multi-local operator. This is an
identity that exactly holds on the lattice.

\subsection{$X_A(x)$ in the continuum limit}

We next consider, on the basis of the perturbation theory, how the $U(1)_A$
breaking term~$X_A(x)$ in~Eq.~\eqref{eq:(3.6)} behaves in the quantum continuum
limit. Since $X_A(x)$ is a dimension~$4$ operator that is proportional to the
lattice spacing~$a$ (because it arises from lattice artifacts), we set
\begin{equation}
   X_A(x)\equiv a\mathcal{O}_5(x),
\label{eq:(3.7)}
\end{equation}
where $\mathcal{O}_5(x)$ is a dimension~$5$ operator. Then, from dimension
counting and the covariance under lattice symmetries~\eqref{eq:(3.5)},
subtractions which are required to define a renormalized composite
operator~$\mathcal{O}_5^R(x)$ in the continuum limit are given
by~\cite{Bochicchio:1985xa} (see also Ref.~\cite{Testa:1998ez}),
\begin{align}
   \mathcal{O}_5^R(x)
   &=\mathcal{Z}_5\biggl\{
   \mathcal{O}_5(x)
   +\frac{1}{a}(\mathcal{Z}_A-1)\partial_\mu\Breve{\jmath}_{5\mu}(x)
   +\frac{1}{a}\mathcal{Z}_{F\widetilde F}
   \epsilon_{\mu\nu\rho\sigma}\tr\left[F_{\mu\nu}(x)F_{\rho\sigma}(x)\right]
\notag\\
   &\qquad\qquad{}
   +\frac{1}{a^2}\mathcal{Z}_P\Breve{P}(x)
   +\frac{1}{a}(\text{dim.\ $4$ BRS non-invariant operators})
   \biggr\}
\notag\\
   &\qquad{}
   +\sum_j\mathcal{Z}_5^{(j)}\mathcal{O}_5^{(j)R}(x),
\label{eq:(3.8)}
\end{align}
where the explicit form of power-subtraction operators is written only for BRS
invariant ones.\footnote{One sees that there is no dimension~$4$ (or less) BRS
invariant operator that contains $c(x)$, $\Bar{c}(x)$ or~$B(x)$ and complies
with property~\eqref{eq:(3.5)}.} As indicated in~Eq.~\eqref{eq:(3.8)},
generally, one might have a mixing with BRS non-invariant operators depending
on the insertion operator~$\mathcal{O}(y,z,\dotsc)$ in~Eq.~\eqref{eq:(3.6)}.
The last line of~Eq.~\eqref{eq:(3.8)} represents a possible mixing with other
(renormalized) dimension~$5$ operators. We have used the continuum theory
language in the above expression ($\Breve{P}(x)$ is the classical continuum
limit of $P(x)$~\eqref{eq:(3.4)}), because the present considerations are
meaningful only in the continuum limit.

Eqs.~\eqref{eq:(3.7)} and~\eqref{eq:(3.8)} show that
\begin{subequations}
\label{eq:(3.9)}
\begin{align}
   X_A(x)&=
   (1-\mathcal{Z}_A)\partial_\mu\Breve{\jmath}_{5\mu}(x)
\label{eq:(3.9a)}\\
   &\qquad{}
   -\mathcal{Z}_{F\widetilde F}
   \epsilon_{\mu\nu\rho\sigma}\tr\left[F_{\mu\nu}(x)F_{\rho\sigma}(x)\right]
\label{eq:(3.9b)}\\
   &\qquad{}
   -\frac{1}{a}\mathcal{Z}_P\Breve{P}(x)
\label{eq:(3.9c)}\\
   &\qquad{}
   +(\text{dim.\ $4$ BRS non-invariant operators})
\label{eq:(3.9d)}\\
   &\qquad{}
   +a\mathcal{Z}_5^{-1}
   \left[\mathcal{O}_5^R(x)
   -\sum_j\mathcal{Z}_5^{(j)}\mathcal{O}_5^{(j)R}(x)\right],
\label{eq:(3.9e)}
\end{align}
\end{subequations}
and the lattice identity~\eqref{eq:(3.6)} then reads
\begin{subequations}
\label{eq:(3.10)}
\begin{align}
   &\mathcal{Z}_A\left\langle
   \partial_\mu\Breve{\jmath}_{5\mu}(x)\mathcal{O}(y,z,\dotsc)
   \right\rangle
\label{eq:(3.10a)}\\
   &=-\mathcal{Z}_{F\widetilde F}\left\langle
   \epsilon_{\mu\nu\rho\sigma}\tr\left[F_{\mu\nu}(x)F_{\rho\sigma}(x)\right]
   \mathcal{O}(y,z,\dotsc)\right\rangle
\label{eq:(3.10b)}\\
   &\qquad{}
   +2\left(M-\frac{1}{2a}\mathcal{Z}_P\right)
   \left\langle\Breve{P}(x)\mathcal{O}(y,z,\dotsc)\right\rangle
\label{eq:(3.10c)}\\
   &\qquad{}
   +\left\langle(\text{dim.\ $4$ BRS non-invariant operators})
   \mathcal{O}(y,z,\dotsc)\right\rangle
\label{eq:(3.10d)}\\
   &\qquad{}
   +\left\langle
   a\mathcal{Z}_5^{-1}\left[
   \mathcal{O}_5^R(x)-
   \sum_j\mathcal{Z}_5^{(j)}\mathcal{O}_5^{(j)R}(x)\right]
   \mathcal{O}(y,z,\dotsc)\right\rangle
\label{eq:(3.10e)}\\
   &\qquad{}
   +i\left\langle\frac{1}{a^4}\frac{\partial}{\partial\theta(x)}
   \delta_\theta\mathcal{O}(y,z,\dotsc)\right\rangle.
\label{eq:(3.10f)}
\end{align}
\end{subequations}

Now, for simplicity, we assume \emph{temporarily\/} that the inserted
operator~$\mathcal{O}(y,z,\dotsc)$ is gauge (or BRS) invariant and
renormalizable without subtracting power-divergent terms. That is,
\begin{equation}
   \mathcal{O}^R(y,z,\dotsc)
   =\mathcal{Z}_{\mathcal{O}}\mathcal{O}(y,z,\dotsc)
   +\sum_j\mathcal{Z}^{(j)}\mathcal{O}^{(j)R}(y,z,\dotsc),
\label{eq:(3.11)}
\end{equation}
where all renormalization constants are at most logarithmically divergent. We
further assume that all the points~$x$, $y$, $z$, \dots,
in~Eq.~\eqref{eq:(3.10)} differ from each other. In this situation, the contact
term~\eqref{eq:(3.10f)} vanishes. The contribution of the dimension~$5$
operators~\eqref{eq:(3.10e)} also vanishes in the continuum limit because of
the overall factor of~$a$ (when all the points~$x$, $y$, $z$, \dots, differ, no
$O(1/a)$ ultraviolet divergence arises that can cancel the overall factor~$a$).
There is no mixing of~$X_A(x)$ with BRS non-invariant operators because the
inserted operator~$\mathcal{O}(y,z,\dotsc)$ is BRS invariant. Therefore, in
this assumed situation, we have
\begin{align}
   \mathcal{Z}_A\left\langle
   \partial_\mu\Breve{\jmath}_{5\mu}(x)\mathcal{O}(y,z,\dotsc)
   \right\rangle
   &=-\mathcal{Z}_{F\widetilde F}\left\langle
   \epsilon_{\mu\nu\rho\sigma}\tr\left[F_{\mu\nu}(x)F_{\rho\sigma}(x)\right]
   \mathcal{O}(y,z,\dotsc)\right\rangle
\notag\\
   &\qquad{}
   +2\left(M-\frac{1}{2a}\mathcal{Z}_P\right)
   \left\langle\Breve{P}(x)\mathcal{O}(y,z,\dotsc)\right\rangle.
\label{eq:(3.12)}
\end{align}
From this (anomalous) conservation law of the axial $U(1)$ current, one might
infer that for an infinitesimal lattice spacing the $U(1)_A$ symmetry broken by
the lattice regularization is restored by tuning the bare mass parameter~$M$
such that
\begin{equation}
   M-\frac{1}{2a}\mathcal{Z}_P\to0.
\label{eq:(3.13)}
\end{equation}
We call this a chiral symmetric
limit~\cite{Karsten:1980wd,Bochicchio:1985xa,Curci:1986sm}.

Incidentally, as for the domain-wall fermion~\cite{Kaplan:1992bt,Shamir:1993zy}
and for the overlap fermion~\cite{Neuberger:1997fp,Neuberger:1998wv}, if the
lattice Dirac operator~$D$ satisfies the GW
relation~$\gamma_5D+D\gamma_5=aD\gamma_5D$~\cite{Ginsparg:1981bj}, it is
possible to define $j_{5\mu}(x)$ in~Eq.~\eqref{eq:(3.2)} such that
$X_A(x)=a\mathcal{O}_5(x)=a\tr[\Bar{\psi}(x)D\gamma_5D\psi(x)]$. Suppose
that~$M=0$. Then, in the correlation function~\eqref{eq:(3.6)},
contractions of gluino fields in this~$\mathcal{O}_5(x)$ and gluino fields in
the inserted operator~$\mathcal{O}(y,z,\dotsc)$ produce a term being
proportional to~$\delta_{x,y}\delta_{x,z}$, \dots, which does not contribute to
the divergence of the operator~$\mathcal{O}_5(x)$. On the other hand, the
contraction of gluino fields within~$\mathcal{O}_5(x)$ produces the combination
\begin{equation}
   \mathcal{O}_5(x)\to\frac{1}{a}\frac{1}{a^4}
   \frac{1}{2}\tr_D\left[\Hat{\gamma_5}^{aa}(x,x)\right],
\label{eq:(3.14)}
\end{equation}
where $\tr_D$ is the trace over Dirac indices. The modified chiral matrix in
this expression is defined by~\cite{Luscher:1998kn,Niedermayer:1998bi}
\begin{equation}
   \Hat{\gamma_5}^{ab}(x,y)
   \equiv\gamma_5
   2\tr\left[T^a\left(1-aD\right)T^b\right]\delta_{x,y}.
\label{eq:(3.15)}
\end{equation}
From the GW relation, one has the identity
$\sum_{z,c}\Hat{\gamma_5}^{ac}(x,z)\Hat{\gamma_5}^{cb}(z,y)
=\delta^{ab}\delta_{x,y}$ and, from this~\cite{Luscher:1998kn},
\begin{equation}
   \sum_x\frac{1}{2}\tr_D\left[\delta\Hat{\gamma_5}^{aa}(x,x)\right]=0,
\label{eq:(3.16)}
\end{equation}
where $\delta$~denotes an arbitrary infinitesimal local variation of the gauge
field. This relation implies that the vertices that result from the composite
operator~\eqref{eq:(3.14)} identically vanish when the momentum being conjugate
to the position~$x$ vanishes. In possible subtractions~\eqref{eq:(3.8)}, this
property is shared by operators $\partial_\mu\Breve{\jmath}_{5\mu}(x)$
and~$\epsilon_{\mu\nu\rho\sigma}\tr\left[F_{\mu\nu}(x)F_{\rho\sigma}(x)\right]$,
but not by~$\Breve{P}(x)$. This shows that $\mathcal{Z}_P=0$
when~$M=0$.\footnote{On the other hand, $\mathcal{Z}_{F\Tilde{F}}$ reproduces
the correct axial
anomaly~\cite{Kikukawa:1998pd,Fujikawa:1998if,Adams:1998eg,Suzuki:1998yz,%
Chiu:1998qv}.} That is, when $D$ satisfies the GW relation, we generally
have\footnote{I am grateful to Yoshio Kikukawa for a clarifying discussion on
this point.}
\begin{equation}
   \mathcal{Z}_P\propto aM.
\label{eq:(3.17)}
\end{equation}
This ensures Eq.~\eqref{eq:(3.13)} for~$M\to0$.

\section{SUSY WT identity on the lattice}
\label{sec:4}

\subsection{Derivation of the SUSY WT identity}

A lattice WT identity associated with the SUSY transformation can also be
derived in a similar way as the above $U(1)_A$ WT identity. The localized
version of the SUSY transformation in the continuum is given by
\begin{align}
   \delta_\xi A_\mu(x)&=\Bar{\xi}(x)\gamma_\mu\psi(x),&&
\notag\\
   \delta_\xi\psi(x)&=-\frac{1}{2}\sigma_{\mu\nu}\xi(x)F_{\mu\nu}(x),&
   \delta_\xi\Bar{\psi}(x)
   &=\frac{1}{2}\Bar{\xi}(x)\sigma_{\mu\nu}F_{\mu\nu}(x),
\label{eq:(4.1)}
\end{align}
where the Grassmann-odd parameters~$\xi(x)$ obey
$\Bar{\xi}(x)=\xi^T(x)(-C^{-1})$. Corresponding to this, we may adopt the
following localized lattice SUSY transformation,
\begin{align}
   \delta_\xi U_\mu(x)&=iag\frac{1}{2}
   \left[
   \Bar{\xi}(x)\gamma_\mu\psi(x)U_\mu(x)
   +\Bar{\xi}(x+a\hat\mu)\gamma_\mu U_\mu(x)\psi(x+a\hat\mu)
   \right],
\notag\\
   \delta_\xi U_\mu^\dagger(x)&=-iag\frac{1}{2}
   \left[
   \Bar{\xi}(x)\gamma_\mu U_\mu^\dagger(x)\psi(x)
   +\Bar{\xi}(x+a\hat\mu)\gamma_\mu\psi(x+a\hat\mu)U_\mu^\dagger(x)
   \right],
\notag\\
   \delta_\xi\psi(x)&=-\frac{1}{2}\sigma_{\mu\nu}\xi(x)P_{\mu\nu}(x),\qquad
   \delta_\xi\Bar{\psi}(x)
   =\frac{1}{2}\Bar{\xi}(x)\sigma_{\mu\nu}P_{\mu\nu}(x),
\label{eq:(4.2)}
\end{align}
where $P_{\mu\nu}(x)$ is the clover plaquette, defined by
\begin{equation}
   P_{\mu\nu}(x)=\frac{1}{4}\sum_{i=1}^4\frac{1}{2ia^2g}
   \left[U_{i\mu\nu}(x)-U_{i\mu\nu}^\dagger(x)\right],
\label{eq:(4.3)}
\end{equation}
and
\begin{align}
   U_{1\mu\nu}(x)
   &\equiv U_\mu(x)U_\nu(x+a\hat\mu)U_\mu^\dagger(x+a\hat\nu)U_\nu^\dagger(x),
\notag\\
   U_{2\mu\nu}(x)
   &\equiv U_\nu(x)U_\mu^\dagger(x-a\hat\mu+a\hat\nu)
   U_\nu^\dagger(x-a\hat\mu)U_\mu(x-a\hat\mu),
\notag\\
   U_{3\mu\nu}(x)
   &\equiv U_\mu^\dagger(x-a\hat\mu)U_\nu^\dagger(x-a\hat\mu-a\hat\nu)
   U_\mu(x-a\hat\mu-a\hat\nu)U_\nu(x-a\hat\nu),
\notag\\
   U_{4\mu\nu}(x)
   &\equiv U_\nu^\dagger(x-a\hat\nu)U_\mu(x-a\hat\nu)
   U_\nu(x+a\hat\mu-a\hat\nu)U_\mu^\dagger(x).
\label{eq:(4.4)}
\end{align}
This definition of~$\delta_\xi$~\cite{Taniguchi:1999fc} is advantageous because
$P_{\mu\nu}(x)$ possesses the same transformation properties as the
continuum~$F_{\mu\nu}(x)$ under~$\mathcal{P}$ and~$\mathcal{T}$. That is,
\begin{equation}
   P_{0k}(x)\xrightarrow{\mathcal{P}}-P_{0k}(x_0,-\Vec{x}),\qquad
   P_{kl}(x)\xrightarrow{\mathcal{P}}P_{kl}(x_0,-\Vec{x}),
\label{eq:(4.5)}
\end{equation}
and
\begin{equation}
   P_{0k}(x)\xrightarrow{\mathcal{T}}-P_{0k}(-x_0,\Vec{x}),\qquad
   P_{kl}(x)\xrightarrow{\mathcal{T}}P_{kl}(-x_0,\Vec{x}).
\label{eq:(4.6)}
\end{equation}

We define that the ghost, anti-ghost and auxiliary fields as singlets under the
above (localized) SUSY transformation~$\delta_\xi$. Then one sees that the BRS
transformation~$s_0$~\eqref{eq:(2.14)} and~$\delta_\xi$ anti-commute with each
other, and as the consequence, the SUSY variation of the gauge fixing and ghost
actions is BRS exact:
\begin{equation}
   \delta_\xi S_{\text{GF}+\text{FP}}^{(0)}
   =-s_0a^4\sum_x2\tr\left[
   \Bar{c}(x)\partial_\mu^*\delta_\xi A_\mu(x)\right].
\label{eq:(4.7)}
\end{equation}
The explicit form of~$\delta_\xi A_\mu(x)$, a transformation of the gauge
potential induced by lattice SUSY transformation~\eqref{eq:(4.2)}, is
\begin{align}
   \delta_\xi A_\mu(x)
   &=\frac{1}{2}
   \Biggl\{
   \Bar{\xi}(x)\gamma_\mu
   \frac{iag\varDelta_{A_\mu(x)}}{\exp\left[iag\varDelta_{A_\mu(x)}\right]-1}
   \psi(x)
\notag\\
   &\qquad\qquad{}
   +\Bar{\xi}(x+a\Hat{\mu})\gamma_\mu
   \frac{iag\varDelta_{A_\mu(x)}}{1-\exp\left[-iag\varDelta_{A_\mu(x)}\right]}
   \psi(x+a\hat{\mu})
   \biggr\}.
\label{eq:(4.8)}
\end{align}

Now, we consider the variation of the lattice action~\eqref{eq:(2.17)}
under~Eq.~\eqref{eq:(4.2)}. As the general structure of the variation (noting
Eq.~\eqref{eq:(4.7)}), we have
\begin{align}
   \delta_\xi S^{(0)}&\equiv a^4\sum_x\Bar{\xi}(x)
   \Bigg\{-\partial_\mu^*S_\mu(x)+M\chi(x)
\notag\\
   &\qquad\qquad\qquad{}
   -s_0\sum_y2\tr\left[\Bar{c}(y)\partial_\mu^{*y}
   \frac{\partial}{\partial\Bar{\xi}(x)}\delta_\xi A_\mu(y)\right]
   +X_S(x)\biggr\}.
\label{eq:(4.9)}
\end{align}
Here, $\chi(x)$ is a gauge invariant fermionic field, which arises from the
variation of the gluino mass term~$S_{\text{mass}}^{(0)}$,
\begin{equation}
   \chi(x)\equiv\sigma_{\mu\nu}\tr\left[\psi(x)P_{\mu\nu}(x)\right]
   \xrightarrow{a\to0}
   \Breve{\chi}(x)\equiv\sigma_{\mu\nu}\tr\left[\psi(x)F_{\mu\nu}(x)\right].
\label{eq:(4.10)}
\end{equation}
In~Eq.~\eqref{eq:(4.9)}, we defined a lattice supercurrent~$S_\mu(x)$, whose
classical continuum limit is
\begin{equation}
   S_\mu(x)\xrightarrow{a\to0}
   \Breve{S}_\mu(x)\equiv
   -\sigma_{\rho\sigma}\gamma_\mu
   \tr\left[\psi(x)F_{\rho\sigma}(x)\right].
\label{eq:(4.11)}
\end{equation}
However, in~Eq.~\eqref{eq:(4.9)}, the separation between
$-\partial_\mu^*S_\mu(x)$ and~$X_S(x)$, a SUSY breaking associated with the
lattice regularization, is not unique even under condition~\eqref{eq:(4.11)};
there remains an $O(a)$~ambiguity in the definition of~$X_S(x)$. We thus
require that the breaking~$X_S(x)$ is a gauge-invariant local combination of
lattice fields which behaves in an identical way as~$\chi(x)$~\eqref{eq:(4.10)}
under lattice discrete transformations. In particular, we require that
\begin{equation}
   X_S(x_0,\Vec{x})\xrightarrow{\mathcal{P}}i\gamma_0X_S(x_0,-\Vec{x}),\qquad
   X_S(x_0,\Vec{x})\xrightarrow{\mathcal{T}}i\gamma_0\gamma_5X_S(-x_0,\Vec{x}).
\label{eq:(4.12)}
\end{equation}
Then, for any multi-local operator~$\mathcal{O}(y,z,\dotsc)$, we have an exact
identity on the lattice,
\begin{align}
   &\left\langle\partial_\mu^*S_\mu(x)\mathcal{O}(y,z,\dotsc)\right\rangle
\notag\\
   &=\left\langle\left[M\chi(x)+ X_S(x)\right]\mathcal{O}(y,z,\dotsc)
   \right\rangle
\notag\\
   &\qquad{}
   -\left\langle
   s_0\sum_w2\tr\left[\Bar{c}(w)\partial_\mu^{*w}
   \frac{\partial}{\partial\Bar{\xi}(x)}\delta_\xi A_\mu(w)\right]
   \mathcal{O}(y,z,\dotsc)\right\rangle
\notag\\
   &\qquad{}
   -\left\langle
   \frac{1}{a^4}\frac{\partial}{\partial\Bar{\xi}(x)}\delta_\xi
   \mathcal{O}(y,z,\dotsc)\right\rangle.
\label{eq:(4.13)}
\end{align}

\subsection{$X_S(x)$ in the continuum limit}

We next investigate, within perturbation theory, how the SUSY breaking
term~$X_S(x)$ in~Eq.~\eqref{eq:(4.13)} behaves in the continuum limit. $X_S(x)$
is a dimension~$9/2$ operator that is proportional to the lattice spacing~$a$
(because it results from the lattice regularization). Thus we set
\begin{equation}
   X_S(x)=a\mathcal{O}_{11/2}(x),
\label{eq:(4.14)}
\end{equation}
where $\mathcal{O}_{11/2}(x)$ is a dimension~$11/2$ operator. To define a
renormalized finite operator~$\mathcal{O}_{11/2}^R(x)$ in the continuum limit,
we generally need the subtraction of operators of mass-dimension $9/2$ or less
with power-diverging coefficients. Those operators must, by our assumption,
behave identically to~$\chi(x)$~\eqref{eq:(4.10)} under lattice discrete
transformations (as~Eq.~\eqref{eq:(4.12)}). A gauge (or BRS) invariant class of
such operators, without containing the ghost, anti-ghost and auxiliary fields,
are enumerated in~Appendix~B of~Ref.~\cite{Farchioni:2001wx}. Utilizing this
result, we have\footnote{Here, we have noted the Bianchi identity,
$\epsilon_{\mu\nu\rho\sigma}D_\nu F_{\rho\sigma}(x)=0$.}
\begin{align}
   &\mathcal{O}_{11/2}^R(x)
\notag\\
   &=\mathcal{Z}_{11/2}\biggl\{
   \mathcal{O}_{11/2}(x)+\frac{1}{a}(\mathcal{Z}_S-1)\partial_\mu\Breve{S}_\mu(x)
   +\frac{1}{a}\mathcal{Z}_T\partial_\mu\Breve{T}_\mu(x)
\notag\\
   &\qquad\qquad{}
   +\frac{1}{a^2}\mathcal{Z}_\chi\Breve{\chi}(x)
\notag\\
   &\qquad\qquad{}
   +\frac{1}{a}\mathcal{Z}_{3F}\tr\left[\psi(x)\Bar{\psi}(x)\psi(x)\right]
\notag\\
   &\qquad\qquad{}
   +\frac{1}{a}\mathcal{Z}_{\text{EOM}}
   \left\{\gamma_\mu\tr\left[\psi(x)D_\nu F_{\mu\nu}(x)\right]
   -s_0\gamma_\mu\tr\left[\psi(x)\partial_\mu\Bar{c}(x)\right]\right\}
\notag\\
   &\qquad\qquad{}
   +\frac{1}{a}(\text{dim. 9/2 BRS invariant operators containing $c$, $\Bar{c}$ or $B$})
\notag\\
   &\qquad\qquad{}
   +\frac{1}{a}(\text{dim.\ $9/2$ BRS non-invariant operators})
   \biggr\}
\notag\\
   &\qquad{}
   +\sum_j\mathcal{Z}_{11/2}^{(j)}\mathcal{O}_{11/2}^{(j)R}(x),
\label{eq:(4.15)}
\end{align}
where
\begin{equation}
   \Breve{T}_\mu(x)\equiv2\gamma_\nu
   \tr\left[\psi(x)F_{\mu\nu}(x)\right].
\label{eq:(4.16)}
\end{equation}
In the above expression, the last line represents possible mixing with other
(renormalized) dimension~$11/2$ operators. From this and Eq.~\eqref{eq:(4.14)},
we have
\begin{subequations}
\label{eq:(4.17)}
\begin{align}
   X_S(x)&=
   (1-\mathcal{Z}_S)\partial_\mu\Breve{S}_\mu(x)
\label{eq:(4.17a)}\\
   &\qquad{}
   -\mathcal{Z}_T\partial_\mu\Breve{T}_\mu(x)
\label{eq:(4.17b)}\\
   &\qquad{}
   -\frac{1}{a}\mathcal{Z}_\chi\Breve{\chi}(x)
\label{eq:(4.17c)}\\
   &\qquad{}
   -\mathcal{Z}_{3F}\tr\left[\psi(x)\Bar{\psi}(x)\psi(x)\right]
\label{eq:(4.17d)}\\
   &\qquad{}
   -\mathcal{Z}_{\text{EOM}}
   \left\{\gamma_\mu\tr\left[\psi(x)D_\nu F_{\mu\nu}(x)\right]
   -s_0\gamma_\mu\tr\left[\psi(x)\partial_\mu\Bar{c}(x)\right]\right\}
\label{eq:(4.17e)}\\
   &\qquad{}
   +(\text{dim. 9/2 BRS invariant operators containing $c$, $\Bar{c}$ or $B$})
\label{eq:(4.17f)}\\
   &\qquad{}
   +(\text{dim.\ $9/2$ BRS non-invariant operators})
\label{eq:(4.17g)}\\
   &\qquad{}
   +a\mathcal{Z}_{11/2}^{-1}\left[
   \mathcal{O}_{11/2}^R(x)
   -\sum_j\mathcal{Z}_{11/2}^{(j)}\mathcal{O}_{11/2}^{(j)R}(x)
   \right],
\label{eq:(4.17h)}
\end{align}
\end{subequations}
and thus, combined with~Eq.~\eqref{eq:(4.13)},
\begin{subequations}
\label{eq:(4.18)}
\begin{align}
   &\left\langle
   \left[\mathcal{Z}_S\partial_\mu\Breve{S}_\mu(x)
   +\mathcal{Z}_T\partial_\mu\Breve{T}_\mu(x)
   \right]\mathcal{O}(y,z,\dotsc)
   \right\rangle
\label{eq:(4.18a)}\\
   &=\left(M-\frac{1}{a}\mathcal{Z}_\chi\right)
   \left\langle\Breve{\chi}(x)\mathcal{O}(y,z,\dotsc)\right\rangle
\label{eq:(4.18b)}\\
   &\qquad{}
   -\mathcal{Z}_{3F}
   \left\langle\tr\left[\psi(x)\Bar{\psi}(x)\psi(x)\right]
   \mathcal{O}(y,z,\dotsc)\right\rangle
\label{eq:(4.18c)}\\
   &\qquad{}
   -\mathcal{Z}_{\text{EOM}}\left\langle
   \frac{1}{2}
   \gamma_\mu\psi^a(x)\frac{\delta\Breve{S}_{\text{tot}}^{(0)}}{\delta A_\mu^a(x)}
   \mathcal{O}(y,z,\dotsc)\right\rangle
\label{eq:(4.18d)}\\
   &\qquad{}
   +\left\langle
   (\text{dim. 9/2 BRS inv.\ op.\ containing $c$, $\Bar{c}$ or $B$})
   \mathcal{O}(y,z,\dotsc)\right\rangle
\label{eq:(4.18e)}\\
   &\qquad{}
   +\left\langle
   (\text{dim.\ $9/2$ BRS non-inv.\ op.})
   \mathcal{O}(y,z,\dotsc)\right\rangle
\label{eq:(4.18f)}\\
   &\qquad{}
   +a\mathcal{Z}_{11/2}^{-1}\left\langle
   \left[
   \mathcal{O}_{11/2}^R(x)-
   \sum_j\mathcal{Z}_{11/2}^{(j)}\mathcal{O}_{11/2}^{(j)R}(x)\right]
   \mathcal{O}(y,z,\dotsc)\right\rangle
\label{eq:(4.18g)}\\
   &\qquad{}
   -\left\langle
   \frac{1}{a^4}\frac{\partial}{\partial\Bar{\xi}(x)}\delta_\xi
   \mathcal{O}(y,z,\dotsc)\right\rangle,
\label{eq:(4.18h)}
\end{align}
\end{subequations}
where we have expressed operator~\eqref{eq:(4.17e)} in terms of the variations
of the total action of the continuum theory,
$\Breve{S}_{\text{tot}}^{(0)}$~\eqref{eq:(2.18)},\footnote{Although
in~Eq.~\eqref{eq:(4.18d)} one has another term,
$-ig\gamma_\mu\tr[\psi(x)\Bar{\psi}(x)\gamma_\mu\psi(x)]$, this combination
identically vanishes as shown in~\ref{sec:A}.} and absorbed the $s_0$-exact
term in~Eq.~\eqref{eq:(4.13)} into~Eq.~\eqref{eq:(4.18e)}. As shown
in~Eqs.~\eqref{eq:(4.17g)} and~\eqref{eq:(4.18f)}, generally, we might have
mixing with BRS non-invariant operators depending on the insertion
operator~$\mathcal{O}(y,z,\dotsc)$; this mixing has actually been observed in a
one-loop calculation~\cite{Taniguchi:1999fc}.

In Eq.~\eqref{eq:(4.15)}, and consequently in~Eqs.~\eqref{eq:(4.17d)}
and~\eqref{eq:(4.18c)}, we have an operator
\begin{equation}
  \mathcal{O}_S(x)\equiv\tr\left[\psi(x)\Bar{\psi}(x)\psi(x)\right],
\label{eq:(4.19)}
\end{equation}
that is cubic in the fermion field.
In~Refs.~\cite{Curci:1986sm,Farchioni:2001wx}, on the other hand, one does
not encounter such a mixing with a three-fermion operator.
In~Ref.~\cite{Curci:1986sm}, this operator was not noticed at all, while
in~Ref.~\cite{Farchioni:2001wx}, it seems that only the case of the gauge group
$SU(2)$ is considered for which
this three-fermion operator identically vanishes $\mathcal{O}_S(x)\equiv0$.
The fact is that, as we analyse in detail in~\ref{sec:A},
$\mathcal{O}_S$ does not generally vanish, and since $\mathcal{O}_S(x)$
and~$\chi(x)$ transform in a completely identical manner under lattice
symmetries,\footnote{Since they rotate in opposite angles under $U(1)_A$, if
the $U(1)_A$ symmetry were exactly preserved, one could get rid of the
possibility of~$\mathcal{O}_S(x)$. The lattice regularization inevitably
breaks the $U(1)_A$, however, to reproduce the axial anomaly.} we cannot
exclude $\mathcal{O}_S(x)$ from the operator renormalization by a simple
symmetry argument alone.

For simplicity, as we did in~Eq.~\eqref{eq:(3.12)}, let us \emph{temporarily\/}
assume that the inserted operator~$\mathcal{O}(y,z,\dotsc)$
in~Eq.~\eqref{eq:(4.18)} is BRS invariant and renormalizable without
subtracting power-divergent terms as~Eq.~\eqref{eq:(3.11)}; we assume also that
all the points~$x$, $y$, $z$, \dots, differ. Then, by a similar argument used
to obtain Eq.~\eqref{eq:(3.12)}, we have in the continuum limit
\begin{align}
   &\left\langle
   \left[\mathcal{Z}_S\partial_\mu\Breve{S}_\mu(x)
   +\mathcal{Z}_T\partial_\mu\Breve{T}_\mu(x)
   \right]\mathcal{O}(y,z,\dotsc)
   \right\rangle
\notag\\
   &=\left(M-\frac{1}{a}\mathcal{Z}_\chi\right)
   \left\langle\Breve{\chi}(x)\mathcal{O}(y,z,\dotsc)\right\rangle
\notag\\
   &\qquad{}
   -\mathcal{Z}_{3F}
   \left\langle\tr\left[\psi(x)\Bar{\psi}(x)\psi(x)\right]
   \mathcal{O}(y,z,\dotsc)\right\rangle,
\label{eq:(4.20)}
\end{align}
where we neglected contribution~\eqref{eq:(4.18e)}, because any BRS invariant
operator with the dimension~$9/2$ or less that contains $c(x)$, $\Bar{c}(x)$
or~$B(x)$ is \emph{always\/} BRS exact; the proof is given in~\ref{sec:B}.

After the redefinition of the supercurrent,
$\mathcal{Z}_S\Breve{S}_\mu+\mathcal{Z}_T\Breve{T}_\mu\propto
\Breve{S}_\mu^{\text{new}}$,\footnote{After this redefinition,
$\Breve{S}_\mu^{\text{new}}$ suffers from the superconformal (or gamma-trace)
anomaly~\cite{Abbott:1977in,Curtright:1977cg,Inagaki:1978iu,Majumdar:1980ej,%
Nicolai:1980km,Hagiwara:1979pu,Hagiwara:1980ys,Kumar:1982ng,Nakayama:1983qt},
because $\gamma_\mu\Breve{T}_\mu(x)\neq0$ while
$\gamma_\mu\Breve{S}_\mu(x)\equiv0$.} Eq.~\eqref{eq:(4.20)} takes the form of
the conservation law of a supercurrent. Then the breaking of SUSY due to the
lattice regularization is characterized by the combination
$M-(1/a)\mathcal{Z}_\chi$. One may remove this unphysical SUSY breaking by
tuning the bare mass parameter~$M$ so that this combination vanishes:
\begin{equation}
   M-\frac{1}{a}\mathcal{Z}_\chi\to0.
\label{eq:(4.21)}
\end{equation}
We call this a supersymmetric limit~\cite{Curci:1986sm} (a non-perturbative
method to impose this condition has been proposed
in~Ref.~\cite{Donini:1997hh}). It is considered that the chiral symmetric
limit~\eqref{eq:(3.13)} coincides with this supersymmetric limit and thus
condition~\eqref{eq:(3.13)} defines a unique supersymmetric
theory~\cite{Curci:1986sm}. For this, we must have
\begin{equation}
   \mathcal{Z}_\chi=\frac{1}{2}\mathcal{Z}_P.
\label{eq:(4.22)}
\end{equation}

Eq.~\eqref{eq:(4.20)} shows that, however, the conservation law of the
supercurrent suffers from an exotic breaking if $\mathcal{Z}_{3F}\neq0$. If
this breaking cannot be removed by a local counter term, this would imply an
exotic SUSY anomaly which certainly what we do not expect in the present
system. Even if this anomaly could be removed by a local counter term, its
presence implies that, as Eq.~\eqref{eq:(4.20)} shows, tuning the mass
parameter~\eqref{eq:(4.21)} alone will not lead to a supersymmetric theory,
contradicting with what is believed. Thus we have to ensure that
\begin{equation}
  \mathcal{Z}_{3F}=0.
\label{eq:(4.23)}
\end{equation}
As noted above, however, it is not possible to conclude $\mathcal{Z}_{3F}=0$ by
a simple symmetry argument alone. It appears that one needs some powerful
machinery such as the WZ consistency condition~\cite{Wess:1971yu}. The rest of
this paper will be entirely devoted to the construction of the required
consistency condition and its application to the proof
of~Eqs.~\eqref{eq:(4.22)} and~\eqref{eq:(4.23)}.

\section{Generalized BRS transformation in the lattice theory}
\label{sec:5}

\subsection{General framework}

We want to formulate a certain WZ consistency condition that constrains the
quantum continuum limit of symmetry breaking terms attributed to the lattice
regularization. Because of several reasons, unfortunately, this is a somewhat
complicated task.

First, the algebraic structure of symmetry transformations is rather involved
even in the continuum target theory. In 4D $\mathcal{N}=1$ SYM in the WZ gauge,
the algebra of SUSY transformations contains field-dependent gauge
transformations. Moreover, the algebra closes only under the equation of motion
of the (massless) gluino (i.e., on-shell closure). Thus, to construct a
BRS-like nilpotent operation that contains SUSY (which will be a building block
in the WZ consistency condition), one has to treat the SUSY, translation and
gauge transformations on an equal footing. For this, we adopt the generalized
BRS transformation developed in the continuum 4D $\mathcal{N}=1$ SYM in the WZ
gauge. See Refs.~\cite{White:1992ai,Maggiore:1995gr,Maggiore:1996gg} and
references therein. The on-shell closure can also be incorporated into this
framework by introducing a term that is quadratic in the ``anti-field'' of the
gluino, $K_\psi$ (see, for example, Refs.~\cite{Howe:1990pz,Maggiore:1995gr}).

Second, we have a bare-mass term of the gluino that is inevitable for the
tuning in lattice formulations; this term however explicitly breaks SUSY and
the $U(1)_A$ symmetry. If one wants to constrain the structure of radiative
corrections by using the Slavnov--Taylor (ST) identity or the Zinn-Justin
equation, those relations should not contain tree-level symmetry-breakings such
as the gluino mass term. To circumvent this point, we introduce generalized BRS
doublet fields (spurions) $(u_V,v_V)$ and~$(u_A,v_A)$ which make the mass term
``formally'' BRS invariant~\cite{Blasi:1995vt,Maggiore:1996gg}.

Finally, of course, SUSY and the infinitesimal translation transformations are
not properly realized on lattice variables and, as the consequence, the
nilpotency of the generalized BRS transformation on some of lattice variables
is broken by~$O(a)$. Then we have to carefully separate those $O(a)$ breakings
from the main part of the ST identity or the Zinn-Justin equation.

We thus define a generalized BRS transformation~$s$ in our lattice system
by
\begin{align}
   sA_\mu(x)&\equiv
   \left[D_\mu c\right]^L(x)
   +\left[\Bar{\xi}\gamma_\mu\psi\right]^L(x)
   -it_\nu\partial_\nu^SA_\mu(x),
\notag\\
   s\psi(x)&\equiv
   -ig\{c(x),\psi(x)\}-\frac{1}{2}\sigma_{\mu\nu}\xi P_{\mu\nu}(x)
   -it_\mu\partial_\mu^S\psi(x)+i\theta\gamma_5\psi(x),
\notag\\
   s\Bar{\psi}(x)&=
   -ig\{c(x),\Bar{\psi}(x)\}+\frac{1}{2}\Bar{\xi}\sigma_{\mu\nu}P_{\mu\nu}(x)
   -it_\mu\partial_\mu^S\Bar{\psi}(x)+i\theta\Bar{\psi}(x)\gamma_5,
\notag\\
   sc(x)&\equiv
   -igc(x)^2
   +\Bar{\xi}\gamma_\mu\xi
   \frac{1}{2}\left[A_\mu(x)+A_\mu(x-a\Hat{\mu})\right]
   -it_\mu\partial_\mu^Sc(x),
\notag\\
   s\Bar{c}(x)&\equiv
   B(x)-it_\mu\partial_\mu^S\Bar{c}(x),
\notag\\
   sB(x)&\equiv
   \Bar{\xi}\gamma_\mu\xi\partial_\mu^S\Bar{c}(x)-it_\mu\partial_\mu^SB(x),
\notag\\
   s\xi&\equiv
   i\theta\gamma_5\xi,\qquad
   s\Bar{\xi}=i\theta\Bar{\xi}\gamma_5,
\notag\\
   st_\mu&\equiv
   -i\Bar{\xi}\gamma_\mu\xi,
\notag\\
   s\theta&\equiv0,
\label{eq:(5.1)}
\end{align}
where $\partial_\mu^S$ denotes the symmetric difference operator
\begin{equation}
   \partial_\mu^S\equiv\frac{1}{2}\left(\partial_\mu+\partial_\mu^*\right).
\label{eq:(5.2)}
\end{equation}
In the above expressions, the gauge-ghost~$c(x)$, anti-ghost $\Bar{c}(x)$ and
the auxiliary field~$B(x)$ are common to the conventional lattice BRS
transformation~$s_0$~\eqref{eq:(2.14)}, while $\xi$, $t_\mu$ and~$\theta$ are
newly-introduced \emph{constant\/} ghosts associated with the SUSY, translation
and $U(1)_A$ transformations, respectively.\footnote{For these ghost variables,
we use the same symbols as the corresponding classical transformation
parameters; we think that no confusion will arise.} These constant ghosts
possess \emph{opposite\/} Grassmann parity to the original transformation
parameters; thus, $\xi$ is Grassmann-even, and $t_\mu$ and~$\theta$ are
Grassmann-odd. The constant Grassmann-even ghost~$\xi$ is subject to the
Majorana constraint,\footnote{It may seem strange that the generalized BRS
transformation~$s$~\eqref{eq:(5.1)} transforms $c(x)$ and~$B(x)$ with the SUSY
ghost~$\xi$, though we assumed in~Sec.~\ref{sec:4} that these fields are
singlets under the SUSY transformation. There is no contradiction, however,
because the combination~$\Bar{\xi}\gamma_\mu\xi$ \emph{identically vanishes\/}
when $\xi$ is Grassmann-odd as the original SUSY parameter is. On the other
hand, with such a combination with a Grassmann-even~$\xi$, $s$ in the continuum
limit becomes nilpotent (up to the equation of motion of the gluino).}
\begin{equation}
   \Bar{\xi}=\xi^T(-C^{-1}).
\label{eq:(5.3)}
\end{equation}
Some useful identities that hold for such a Grassmann-even spinor are
summarized in~\ref{sec:C}. In the first relation of~Eq.~\eqref{eq:(5.1)},
$\left[D_\mu c\right]^L(x)$ is given in~Eq.~\eqref{eq:(2.15)}
and~$[\Bar{\xi}\gamma_\mu\psi]^L(x)$ is defined by
\begin{align}
   &\left[\Bar{\xi}\gamma_\mu\psi\right]^L(x)
\notag\\
   &\equiv\frac{1}{2}\Bar{\xi}\gamma_\mu
   \left\{
   \frac{iag\varDelta_{A_\mu(x)}}{\exp\left[iag\varDelta_{A_\mu(x)}\right]-1}
   \psi(x)
   +\frac{iag\varDelta_{A_\mu(x)}}{1-\exp\left[-iag\varDelta_{A_\mu(x)}\right]}
   \psi(x+a\hat{\mu})
   \right\}.
\label{eq:(5.4)}
\end{align}

If there were only the gauge-ghost~$c(x)$ and the $U(1)_A$ ghost~$\theta$
in~Eq.~\eqref{eq:(5.1)}, we would simply have $s^2=0$ on all lattice variables,
because algebras of the gauge and of the $U(1)_A$ close on lattice variables.
However, since the SUSY algebra (that contains the infinitesimal translation)
is not properly realized on lattice variables, the nilpotency of~$s$ is broken
by lattice artifacts. That is, on lattice variables, we have
\begin{subequations}
\label{eq:(5.5)}
\begin{align}
   s^2A_\mu(x)&=O(a),
\label{eq:(5.5a)}\\
   s^2\psi(x)
   &=-\Bar{\xi}\gamma_\mu\xi D_\mu\psi(x)
   +\sigma_{\mu\nu}\xi\Bar{\xi}\gamma_\mu D_\nu\psi(x)+O(a)
\notag\\
   &=\gamma_5\xi\Bar{\xi}\gamma_5\Slash{D}\psi(x)+O(a)
\notag\\
   &=\gamma_5\xi\Bar{\xi}\gamma_5D\psi(x)+O(a),
\label{eq:(5.5b)}\\
   s^2c(x)&=O(a),
\label{eq:(5.5c)}\\
   s^2\Bar{c}(x)&=s^2B(x)=s^2\xi=s^2t_\mu=s^2\theta=0.
\label{eq:(5.5d)}
\end{align}
\end{subequations}
Moreover, as the right-hand side of~Eq.~\eqref{eq:(5.5b)} shows, $s^2=0$ on the
gluino field~$\psi(x)$ holds only under the equation of motion of the massless
gluino, even in the continuum
theory~\cite{White:1992ai,Maggiore:1995gr,Maggiore:1996gg}.
(From the first line to the second line in~Eq.~\eqref{eq:(5.5b)}, we used the
Fierz theorem in~Eq.~\eqref{eq:(A.1)}.)

We define the gauge-fixing term and the ghost-anti-ghost action with respect to
the generalized BRS transformation~$s$ by
\begin{equation}
   S_{\text{GF}+\text{FP}}\equiv-sa^4\sum_x2\tr\left\{
   \Bar{c}(x)\left[\partial_\mu^*A_\mu(x)+\frac{\alpha}{2}B(x)\right]
   \right\}.
\label{eq:(5.6)}
\end{equation}
Note that this $S_{\text{GF}+\text{FP}}$ reduces to our previous one
$S_{\text{GF}+\text{FP}}^{(0)}$~\eqref{eq:(2.13)}, when all new ghost variables,
$\xi$, $t_\mu$ and~$\theta$, vanish.

This is not the end of the story, however. To incorporate the mass term of the
gluino that explicitly breaks SUSY and $U(1)_A$, we must further introduce
$s$-doublet fields $(u_V,v_V)$ and~$(u_A,v_A)$ following the procedure
of~Refs.~\cite{Blasi:1995vt,Maggiore:1996gg}:
\begin{align}
   su_V(x)&\equiv v_V(x)+M-it_\mu\partial_\mu^Su_V(x)-2i\theta u_A(x),
\notag\\
   sv_V(x)&\equiv\Bar{\xi}\gamma_\mu\xi\partial_\mu^Su_V(x)
   -it_\mu\partial_\mu^Sv_V(x)-2i\theta v_A(x),
\notag\\
   su_A(x)&\equiv v_A(x)-it_\mu\partial_\mu^Su_A(x)-2i\theta u_V(x),
\notag\\
   sv_A(x)&\equiv\Bar{\xi}\gamma_\mu\xi\partial_\mu^Su_A(x)
   -it_\mu\partial_\mu^Sv_A(x)-2i\theta\left[v_V(x)+M\right],
\label{eq:(5.7)}
\end{align}
where $u_V(x)$ and~$u_A(x)$ are Grassmann-odd, while $v_V(x)$ and~$v_A(x)$ are
Grassmann-even. On these variables, we have an exact nilpotency of~$s$:
\begin{equation}
      s^2u_V(x)=s^2u_A(x)=s^2v_V(x)=s^2v_A(x)=0.
\label{eq:(5.8)}
\end{equation}
Using these fields, we define a ``generalized'' mass term by
\begin{align}
   S_{\text{mass}}&=-sa^4\sum_x
   \tr\left\{\Bar{\psi}(x)\left[u_V(x)+u_A(x)\gamma_5\right]\psi(x)\right\}
\notag\\
   &=a^4\sum_xM\tr\left[\Bar{\psi}(x)\psi(x)\right]
\notag\\
   &\qquad{}
   +a^4\sum_x\tr\left\{
   \Bar{\psi}(x)\left[v_V(x)+v_A(x)\gamma_5\right]\psi(x)\right\}
\notag\\
   &\qquad{}
   -a^4\sum_x\tr\left\{
   \Bar{\xi}\sigma_{\mu\nu}P_{\mu\nu}(x)
   \left[u_V(x)+u_A(x)\gamma_5\right]\psi(x)\right\}
\notag\\
   &\qquad{}
   -a^4\sum_x(-i)t_\mu\Tilde{\partial}_\mu^S\tr\left\{
   \Bar{\psi}(x)\left[u_V(x)+u_A(x)\gamma_5\right]\psi(x)\right\},
\label{eq:(5.9)}
\end{align}
where $\Tilde{\partial}_\mu^S$ is defined by the Leibniz rule,
\begin{equation}
   \Tilde{\partial}_\mu^S(X_1\dots X_n)
   \equiv(\partial_\mu^SX_1)\dots X_n+\dots+X_1\dots X_{n-1}(\partial_\mu^SX_n).
\label{eq:(5.10)}
\end{equation}
Again, $S_{\text{mass}}$ reduces to our original mass term
$S_{\text{mass}}^{(0)}$~\eqref{eq:(2.6)} when the newly-introduced fields,
$(u_V,v_V)$ and~$(u_A,v_A)$, vanish.

We also introduce source terms for lattice dynamical variables
\begin{align}
   S_{\text{source1}}&\equiv
   a^4\sum_x\bigl[
   J_{A\mu}^a(x)A_\mu^a(x)+\Bar{J}_\psi^a(x)\psi^a(x)+J_c^a(x)c^a(x)
\notag\\
   &\qquad\qquad\qquad\qquad\qquad{}
   +J_{\Bar{c}}^a(x)\Bar{c}^a(x)+J_B^a(x)B^a(x)\bigr]
\label{eq:(5.11)}
\end{align}
(we may define $J_\psi^a(x)$ by~$\Bar{J}_\psi^a(x)\equiv J_\psi^{Ta}(x)(-C^{-1})$)
and, following the standard procedure (see, for example,
Ref.~\cite{Weinberg:1996kr}), source terms associated with the generalized BRS
transformation~$s$,
\begin{align}
   S_{\text{source2}}&\equiv
   a^4\sum_x\left[
   K_{A\mu}^a(x)sA_\mu^a(x)+\Bar{K}_\psi^a(x)s\psi^a(x)+K_c^a(x)sc^a(x)
   \right]
\notag\\
   &\qquad\qquad{}
   -a^4\sum_x\frac{1}{2}
   \Bar{K}_\psi^a(x)\gamma_5\xi\Bar{K}_\psi^a(x)\gamma_5\xi.
\label{eq:(5.12)}
\end{align}
(Again we may define $K_\psi^a(x)$
by~$\Bar{K}_\psi^a(x)\equiv K_\psi^{Ta}(x)(-C^{-1})$.) For the massive theory, it
turns out that we need one more term,
\begin{align}
   S_\Delta&\equiv
   a^4\sum_x
   \Bar{K}_\psi^a(x)\gamma_5\xi
   \Bar{\psi}^a(x)\left[u_V(x)+u_A(x)\gamma_5\right]\gamma_5\xi
\notag\\
   &\qquad{}
   -a^4\sum_x\frac{1}{2}
   \Bar{\psi}^a(x)\left[u_V(x)+u_A(x)\gamma_5\right]\gamma_5\xi
   \Bar{\psi}^a(x)\left[u_V(x)+u_A(x)\gamma_5\right]\gamma_5\xi.
\label{eq:(5.13)}
\end{align}
The last term of~Eq.~\eqref{eq:(5.12)}, which is quadratic in the
source~$K_{\psi}(x)$, and~$S_\Delta$ are required to make the following ST
identity or the Zinn-Justin equation hold off-shell up to $O(a)$ lattice
artifacts.

Thus, our generalized total lattice action is
\begin{equation}
   S_{\text{tot}}\equiv
   S_{\text{gluon}}+S_{\text{gluino}}+S_{\text{GF}+\text{FP}}+S_{\text{mass}}
   +S_{\text{source1}}+S_{\text{source2}}+S_\Delta.
\label{eq:(5.14)}
\end{equation}
Using this, we define the generating functional for the connected diagram~$W$,
\begin{equation}
   e^{-W[J,K,\xi,t,\theta,u,v]}\equiv\int d\mu\,e^{-S_{\text{tot}}},
\label{eq:(5.15)}
\end{equation}
where $d\mu$ denotes the integration measure for dynamical variables,
$U(x,\mu)$, $\psi(x)$, $c(x)$, $\Bar{c}(x)$ and~$B(x)$. An important point to
recognize here is that those newly-introduced variables, $\xi$, $t_\mu$,
$\theta$, $u_V(x)$, $u_A(x)$, $v_V(x)$ and~$v_A(x)$, are all
\emph{non-dynamical}, as the argument of the above~$W$ indicates. One can
always set those external variables zero and then the system reduces to our
original lattice theory in~Sec.~\ref{sec:2}. In this way, the dynamics of the
original system can always be reproduced; yet, those new variables are quite
useful to organize the underlying symmetry structure.

For $\mathcal{P}$ and~$\mathcal{T}$ transformation properties of new variables,
we can set
\begin{align}
   &\xi\xrightarrow{\mathcal{P}}i\gamma_0\xi,\qquad
   \Bar{\xi}\xrightarrow{\mathcal{P}}-i\Bar{\xi}\gamma_0,\qquad
   t_0\xrightarrow{\mathcal{P}}t_0,\qquad
   t_k\xrightarrow{\mathcal{P}}-t_k,\qquad
   \theta\xrightarrow{\mathcal{P}}-\theta,
\notag\\
   &u_V(x_0,\Vec{x})\xrightarrow{\mathcal{P}}u_V(x_0,-\Vec{x}),\qquad
   v_V(x_0,\Vec{x})\xrightarrow{\mathcal{P}}v_V(x_0,-\Vec{x}),
\notag\\
   &u_A(x_0,\Vec{x})\xrightarrow{\mathcal{P}}-u_A(x_0,-\Vec{x}),\qquad
   v_A(x_0,\Vec{x})\xrightarrow{\mathcal{P}}-v_A(x_0,-\Vec{x}),
\label{eq:(5.16)}
\end{align}
and
\begin{align}
   &\xi\xrightarrow{\mathcal{T}}i\gamma_0\gamma_5\xi,\qquad
   \Bar{\xi}\xrightarrow{\mathcal{T}}-i\Bar{\xi}\gamma_5\gamma_0,\qquad
   t_0\xrightarrow{\mathcal{T}}-t_0,\qquad
   t_k\xrightarrow{\mathcal{T}}t_k,\qquad
   \theta\xrightarrow{\mathcal{T}}-\theta,
\notag\\
   &u_V(x_0,\Vec{x})\xrightarrow{\mathcal{T}}u_V(-x_0,\Vec{x}),\qquad
   v_V(x_0,\Vec{x})\xrightarrow{\mathcal{T}}v_V(-x_0,\Vec{x}),
\notag\\
   &u_A(x_0,\Vec{x})\xrightarrow{\mathcal{T}}-u_A(-x_0,\Vec{x}),\qquad
   v_A(x_0,\Vec{x})\xrightarrow{\mathcal{T}}-v_A(-x_0,\Vec{x}),
\label{eq:(5.17)}
\end{align}
so that $s$ in~Eqs.~\eqref{eq:(5.1)} and~\eqref{eq:(5.7)} preserves
transformation properties under $\mathcal{P}$ and~$\mathcal{T}$; for example,
one sees that $[\Bar{\xi}\gamma_\mu\psi]^L(x)$ transforms in an identical way
as~$A_\mu(x)$. Then, it is easy to find appropriate $\mathcal{P}$
and~$\mathcal{T}$ transformations of source fields~$J$ and~$K$ such that the
total action~$S_{\text{tot}}$~\eqref{eq:(5.14)} is invariant under $\mathcal{P}$
and~$\mathcal{T}$.

Now, a crucial property of the above total action~$S_{\text{tot}}$, that can be
obtained by a careful calculation using relations in~\ref{sec:C}, is
\begin{align}
   &sS_{\text{tot}}
   -a^4\sum_x
   \Bar{K}_\psi^{\prime a}(x)\gamma_5\xi
   \Bar{\xi}\gamma_5\frac{\delta}{\delta\Bar{\psi}^a(x)}S_{\text{tot}}
\notag\\
   &=a^4\sum_x
   \Bigl\{
   J_{A_\mu}^a(x)sA_\mu^a(x)
   -\Bar{J}_\psi^a(x)
   \left[s\psi^a(x)-\gamma_5\xi\Bar{K}_\psi^{\prime a}(x)\gamma_5\xi\right]
\notag\\
   &\qquad\qquad\qquad{}
   -J_c^a(x)sc^a(x)
   -J_{\Bar{c}}^a(x)s\Bar{c}^a(x)
   +J_B^a(x)sB^a(x)
   \Bigr\}
\notag\\
   &\qquad{}
   +a^4\sum_x\left[\Bar{\xi}X_S(x)+\theta X_A(x)\right]
   +\Bar{c}\cdot\mathcal{B}_{\Bar{c}}
   +K'\cdot\mathcal{B}_{K'}
   +t\cdot\mathcal{B}_t,
\label{eq:(5.18)}
\end{align}
where
\begin{equation}
   \Bar{K}_\psi'(x)\equiv
   \Bar{K}_\psi(x)
   -\Bar{\psi}(x)\left[u_V(x)+u_A(x)\gamma_5\right].
\label{eq:(5.19)}
\end{equation}
In deriving Eq.~\eqref{eq:(5.18)}, we parametrized the breaking of the super,
$U(1)_A$ and translation symmetries in our original lattice action as
\begin{equation}
   s\left(S_{\text{gluon}}+S_{\text{gluino}}\right)
   \equiv
   a^4\sum_x\left[\Bar{\xi}X_S(x)+i\theta X_A(x)+t_\mu X_{t\mu}(x)\right].
\label{eq:(5.20)}
\end{equation}
This $X_S$ is thus \emph{identical\/} to our previous definition
in~Eq.~\eqref{eq:(4.9)} and $X_A$ is also \emph{identical\/} to~$X_A$
in~Eq.~\eqref{eq:(3.2)}; this correspondence will be the key to our analysis
below. Since the continuum action without the gluon mass term possesses the
super, $U(1)_A$ and translation symmetries, all these breakings, $X_S(x)$,
$X_A(x)$ and~$X_{t\mu}(x)$, are of~$O(a)$.

Explicit forms of the last three combinations in~Eq.~\eqref{eq:(5.18)} are
given by
\begin{equation}
   \Bar{c}\cdot\mathcal{B}_{\Bar{c}}\equiv
   -a^4\sum_x\Bar{c}^a(x)\partial_\mu^*s^2A_\mu^a(x),
\label{eq:(5.21)}
\end{equation}
\begin{align}
   K'\cdot\mathcal{B}_{K'}
   &\equiv
   a^4\sum_x\Bigl\{
   -K_{A_\mu}^a(x)s^2A_\mu^a(x)
   +\Bar{K}_\psi^{\prime a}(x)
   \left[s^2\psi^a(x)-\gamma_5\xi\Bar{\xi}\gamma_5D\psi^a(x)\right]
\notag\\
   &\qquad\qquad\qquad{}
   +K_c^a(x)s^2c^a(x)
   \Bigr\},
\label{eq:(5.22)}
\end{align}
and
\begin{align}
   &t\cdot\mathcal{B}_{t}
\notag\\
   &\equiv
   a^4\sum_x t_\mu X_{t_\mu}(x)
\notag\\
   &\qquad{}
   -a^4\sum_x(-i)t_\mu
   \Bar{K}_\psi^{\prime a}(x)\gamma_5\xi
   \Bigl(\partial_\mu^S\Bar{K}_\psi^a(x)
\notag\\
   &\qquad\qquad\qquad\qquad\qquad\qquad\qquad{}
   -\Tilde{\partial}_\mu^S
   \left\{\Bar{\psi}^a(x)\left[u_V(x)+u_A(x)\gamma_5\right]\right\}
   \Bigr)\gamma_5\xi
\notag\\
   &\qquad{}
   -a^4\sum_x(-i)t_\mu
   \Bar{K}_\psi^{\prime a}(x)\gamma_5\xi
   \left(\partial_\mu^S-\Tilde{\partial}_\mu^S\right)
   \left\{\Bar{\psi}^a(x)\left[u_V(x)+u_A(x)\gamma_5\right]\gamma_5\xi\right\}.
\label{eq:(5.23)}
\end{align}
It is obvious that these three combinations are of~$O(a)$, especially in view
of~Eq.~\eqref{eq:(5.5)}. Another crucial property of~$S_{\text{tot}}$ to note is
\begin{equation}
   \frac{\delta}{\delta\Bar{K}_\psi^a(x)}S_{\text{tot}}
   =s\psi^a(x)-\gamma_5\xi\Bar{K}_\psi^{\prime a}(x)\gamma_5\xi.
\label{eq:(5.24)}
\end{equation}

\subsection{Generalized ST identity or Zinn-Justin equation and an
associated WZ consistency condition}

Now, the combination in the left-hand side of relation~\eqref{eq:(5.18)}
prompts us to consider the following infinitesimal change of variables for the
functional integral~\eqref{eq:(5.15)}. Letting $\varepsilon$ be a Grassmann-odd
parameter,
\begin{equation}
   \Phi(x)\to\Phi(x)+\delta\Phi(x),\qquad\delta\Phi(x)=\varepsilon s\Phi(x),
\label{eq:(5.25)}
\end{equation}
for all variables~$\Phi$ \emph{including\/} external ones, $\xi$, $t_\mu$,
$\theta$, $u_V(x)$ , $u_A(x)$ , $v_V(x)$ and~$v_A(x)$, except the gluino
field~$\psi(x)$, for which,
\begin{equation}
   \psi(x)\to\psi(x)+\delta\psi(x),\qquad
   \delta\psi(x)=\varepsilon
   \left[s\psi(x)-\gamma_5\xi\Bar{K}_\psi'(x)\gamma_5\xi\right].
\label{eq:(5.26)}
\end{equation}
Although the last term depends on~$\psi(x)$ itself through the definition
of~$K_\psi'(x)$~\eqref{eq:(5.19)}, the Jacobian~$J$ associated with this
infinitesimal change of variable is unity, because
\begin{equation}
   \ln J=\varepsilon\sum_x\Bar{\xi}\left[u_V(x)+u_A(x)\gamma_5\right]\xi=0,
\label{eq:(5.27)}
\end{equation}
with the lattice regularization (note Eq.~\eqref{eq:(C.1)}).

From the above change of variables, using~Eq.~\eqref{eq:(5.18)}, we have the
identity
\begin{align}
   &\Biggl\{s\xi\frac{\partial}{\partial\xi}
   +st_\mu\frac{\partial}{\partial t_\mu}
   +s\theta\frac{\partial}{\partial\theta}
\notag\\
   &\qquad{}
   +a^4\sum_x\biggl[
   su_V(x)\frac{\delta}{\delta u_V(x)}
   +su_A(x)\frac{\delta}{\delta u_A(x)}
\notag\\
   &\qquad\qquad\qquad\qquad{}
   +sv_V(x)\frac{\delta}{\delta v_V(x)}
   +sv_A(x)\frac{\delta}{\delta v_A(x)}
   \biggr]
   \Biggr\}W[J,K,\xi,t,\theta,u,v]
\notag\\
   &=\Biggl\langle
   a^4\sum_x
   \Bigl\{
   J_{A_\mu}^a(x)sA_\mu^a(x)
   -\Bar{J}_\psi^a(x)
   \left[s\psi^a(x)-\gamma_5\xi\Bar{K}_\psi^{\prime a}(x)\gamma_5\xi\right]
\notag\\
   &\qquad\qquad\qquad{}
   -J_c^a(x)sc^a(x)
   -J_{\Bar{c}}^a(x)s\Bar{c}^a(x)
   +J_B^a(x)sB^a(x)
   \Bigr\}
\notag\\
   &\qquad\qquad{}
   +a^4\sum_x\left[\Bar{\xi}X_S(x)+i\theta X_A(x)\right]
   +\Bar{c}\cdot\mathcal{B}_{\Bar{c}}
   +K'\cdot\mathcal{B}_{K'}
   +t\cdot\mathcal{B}_t
   \Biggr\rangle_{J,K,\xi,t,\theta,u,v}.
\label{eq:(5.28)}
\end{align}
As usual, we define the 1PI effective action~$\varGamma$ by the Legendre
transform of~$W$ only with respect to the source~$J$:
\begin{align}
   &\varGamma[A_\mu,\psi,c,\Bar{c},B;K,\xi,t,\theta,u,v]
\notag\\
   &\qquad\equiv
   W[J,K,\xi,t,\theta,u,v]
\notag\\
   &\qquad\qquad{}
   -a^4\sum_x\bigl[
   J_{A\mu}^a(x)A_\mu^a(x)+\Bar{J}_\psi^a(x)\psi^a(x)+J_c^a(x)c^a(x)
\notag\\
   &\qquad\qquad\qquad\qquad\qquad{}
   +J_{\Bar{c}}^a(x)\Bar{c}^a(x)+J_B^a(x)B^a(x)\bigr].
\label{eq:(5.29)}
\end{align}
Here and in what follows, we use the same symbols for the integration variables
and the expectation values for notational simplicity; for example, $A_\mu^a(x)$
in the argument of the effective action~$\varGamma$ actually means
$\langle A_\mu^a(x)\rangle_{J,K,\xi,t,\theta,u,v}$.

Then noting relations such as,
\begin{align}
   \frac{\partial}{\partial\xi}W&=\frac{\partial\varGamma}{\partial\xi},
\label{eq:(5.30)}\\
   J_\mu^a(x)
   &=-\frac{\delta\varGamma}{\delta A_\mu^a(x)},
\label{eq:(5.31)}\\
   \left\langle
   s\psi^a(x)-\gamma_5\xi\Bar{K}_\psi^{\prime a}(x)\gamma_5\xi
   \right\rangle_{J,K,\xi,t,\theta,u,v}
   &=\frac{\delta\varGamma}{\delta\Bar{K}_\psi^a(x)}
\label{eq:(5.32)}
\end{align}
(the last one follows from~Eq.~\eqref{eq:(5.24)}) etc., we can express
identity~\eqref{eq:(5.28)} in terms of the effective action~$\varGamma$. This
gives rise to the following identity (the ST identity or the Zinn-Justin
equation in the present context):
\begin{equation}
   \mathcal{S}(\varGamma)
   =\left\langle
   a^4\sum_x\left[\Bar{\xi}X_S(x)+i\theta X_A(x)\right]
   +\Bar{c}\cdot\mathcal{B}_{\Bar{c}}
   +K'\cdot\mathcal{B}_{K'}
   +t\cdot\mathcal{B}_t
   \right\rangle_{J,K,\xi,t,\theta,u,v},
\label{eq:(5.33)}
\end{equation}
where the combination~$\mathcal{S}(F)$ is defined for an arbitrary
functional~$F$ by,
\begin{align}
   \mathcal{S}(F)
   &\equiv
   a^4\sum_x\left[
   \frac{\delta F}{\delta K_{A_\mu}^a(x)}\frac{\delta F}{\delta A_\mu^a(x)}
   +\frac{\delta F}{\delta\Bar{K}_\psi^a(x)}\frac{\delta F}{\delta\psi^a(x)}
   +\frac{\delta F}{\delta K_c^a(x)}\frac{\delta F}{\delta c^a(x)}\right]
\notag\\
   &\qquad{}
   +a^4\sum_x\left[
   s\Bar{c}^a(x)\frac{\delta F}{\delta\Bar{c}^a(x)}
   +sB^a(x)\frac{\delta F}{\delta B^a(x)}\right]
\notag\\
   &\qquad{}
   +s\xi\frac{\partial F}{\partial\xi}
   +st_\mu\frac{\partial F}{\partial t_\mu}
   +s\theta\frac{\partial F}{\partial\theta}
\notag\\
   &\qquad{}
   +a^4\sum_x\biggl[
   su_V(x)\frac{\delta F}{\delta u_V(x)}
   +su_A(x)\frac{\delta F}{\delta u_A(x)}
\notag\\
   &\qquad\qquad\qquad\qquad\qquad{}
   +sv_V(x)\frac{\delta F}{\delta v_V(x)}
   +sv_A(x)\frac{\delta F}{\delta v_A(x)}
   \biggr].
\label{eq:(5.34)}
\end{align}
Corresponding to this combination, we introduce an operation~$\mathcal{D}(F)$
by
\begin{align}
   \mathcal{D}(F)
   &\equiv
   a^4\sum_x\Biggl[
   \frac{\delta F}{\delta A_\mu^a(x)}\frac{\delta}{\delta K_{A_\mu}^a(x)}
   +\frac{\delta F}{\delta K_{A_\mu}^a(x)}\frac{\delta}{\delta A_\mu^a(x)}
\notag\\
   &\qquad\qquad\qquad{}
   +\frac{\delta F}{\delta\Bar{K}_\psi^a(x)}\frac{\delta}{\delta\psi^a(x)}
   +\frac{\delta F}{\delta\psi^a(x)}\frac{\delta}{\delta\Bar{K}_\psi^a(x)}
\notag\\
   &\qquad\qquad\qquad\qquad{}
   +\frac{\delta F}{\delta K_c^a(x)}\frac{\delta}{\delta c^a(x)}
   +\frac{\delta F}{\delta c^a(x)}\frac{\delta}{\delta K_c^a(x)}
   \Biggr]
\notag\\
   &\qquad{}
   +a^4\sum_x\left[
   s\Bar{c}^a(x)\frac{\delta}{\delta\Bar{c}^a(x)}
   +sB^a(x)\frac{\delta}{\delta B^a(x)}\right]
\notag\\
   &\qquad{}
   +s\xi\frac{\partial}{\partial\xi}
   +st_\mu\frac{\partial}{\partial t_\mu}
   +s\theta\frac{\partial}{\partial\theta}
\notag\\
   &\qquad{}
   +a^4\sum_x\biggl[
   su_V(x)\frac{\delta}{\delta u_V(x)}
   +su_A(x)\frac{\delta}{\delta u_A(x)}
\notag\\
   &\qquad\qquad\qquad\qquad\qquad{}
   +sv_V(x)\frac{\delta}{\delta v_V(x)}
   +sv_A(x)\frac{\delta}{\delta v_A(x)}
   \biggr].
\label{eq:(5.35)}
\end{align}
Then, for an \emph{arbitrary\/} (Grassmann-even) functional~$F$, we have
\begin{equation}
   \mathcal{D}(F)\mathcal{S}(F)=0.
\label{eq:(5.36)}
\end{equation}
As one can verify straightforwardly, this relation follows solely from the
nilpotency $s^2=0$ on the variables, $\Bar{c}(x)$, $B(x)$, $\xi$, $t_\mu$,
$\theta$, $u_V(x)$, $u_A(x)$, $v_V(x)$ and~$v_A(x)$; recall
Eqs.~\eqref{eq:(5.5d)} and~\eqref{eq:(5.8)}. Although the nilpotency of~$s$
on the variables, $A_\mu(x)$, $\psi(x)$ and~$c(x)$, is broken by the lattice
regularization, the nilpotency on these variables is not necessary to
derive~Eq.~\eqref{eq:(5.36)}; this is a crucial observation.

Since Eq.~\eqref{eq:(5.36)} holds for arbitrary~$F$, it holds of course
for~$\varGamma$. Then, combined with~Eq.~\eqref{eq:(5.33)},
Eq.~\eqref{eq:(5.36)} provides a strong constraint on the possible form of the
breaking terms, the right-hand side of~Eq.~\eqref{eq:(5.33)}. That is, we have
\begin{equation}
   \mathcal{D}(\varGamma)
   \left\langle
   a^4\sum_x\left[\Bar{\xi}X_S(x)+i\theta X_A(x)\right]
   +\Bar{c}\cdot\mathcal{B}_{\Bar{c}}
   +K'\cdot\mathcal{B}_{K'}
   +t\cdot\mathcal{B}_t
   \right\rangle_{J,K,\xi,t,\theta,u,v}=0.
\label{eq:(5.37)}
\end{equation}

We now expand the effective action~$\varGamma$ in the powers of the
loop-counting parameter~$\hbar$
as~$\varGamma=\varGamma_0+\varGamma_1+\varGamma_2+\dotsb$, where $\varGamma_0$
is the tree level action,
\begin{equation}
   \varGamma_0=S_{\text{tot}}.
\label{eq:(3.38)}
\end{equation}
In the continuum limit, the expectation value in~Eq.~\eqref{eq:(5.37)}
is~$O(\hbar)$ or higher because the $O(a)$ breakings attributed to the lattice
regularization can survive only through radiative corrections. Thus, taking the
$O(\hbar^n)$ ($n\geq1$) term of~Eq.~\eqref{eq:(5.37)}, where $n$ is the lowest
order of the loop expansion in which the expectation value does not vanish, we
have
\begin{align}
   &\mathcal{D}(S_{\text{tot}})
   \left\langle
   a^4\sum_x\left[\Bar{\xi}X_S(x)+i\theta X_A(x)\right]
   +\Bar{c}\cdot\mathcal{B}_{\Bar{c}}
   +K'\cdot\mathcal{B}_{K'}
   +t\cdot\mathcal{B}_t
   \right\rangle_{J,K,\xi,t,\theta,u,v}^{O(\hbar^n)}
\notag\\
   &=0.
\label{eq:(5.39)}
\end{align}
This is the WZ consistency condition that we were seeking.

\section{Simplified consistency condition and its application}
\label{sec:6}

\subsection{Simplified consistency condition}

To simplify the analysis, we consider the consistency
condition~\eqref{eq:(5.39)} for a special configuration of expectation values
and external variables. First, we assume that expectation values, $A_\mu(x)$,
$\psi(x)$ and~$c(x)$, satisfy the equations of motion of the total
action~$S_{\text{tot}}$. This removes the functional derivative with respect
to~$K$ from the expression for~$\mathcal{D}(S_{\text{tot}})$ (see
Eq.~\eqref{eq:(5.35)}) and $\mathcal{D}(S_{\text{tot}})$ simply becomes the
$s$~transformation. We further set
\begin{align}
   &\Bar{c}(x)=B(x)=0,
\label{eq:(6.1)}
\\
   &K_{A_\mu}(x)=K_\psi(x)=K_c(x)=t_\mu=u_V(x)=u_A(x)=v_V(x)=v_A(x)=0.
\label{eq:(6.2)}
\end{align}

We can now relate Eqs.~\eqref{eq:(3.9)} and~\eqref{eq:(4.17)}
to~Eq.~\eqref{eq:(5.39)}. Since we assumed in~Eq.~\eqref{eq:(6.2)}
that all $K$ vanish, expectation values in~Eq.~\eqref{eq:(5.39)} are computed
without the source term for the composite operators,
$S_{\text{source2}}$~\eqref{eq:(5.12)}; we have only the source term for the
elementary fields, $S_{\text{source1}}$~\eqref{eq:(5.11)}. In the language
of~Eqs.~\eqref{eq:(3.6)} and~\eqref{eq:(4.13)}, this situation corresponds to
the situation that the inserted operator~$\mathcal{O}(y,z,\dotsc)$ is a
collection of \emph{elementary fields}. In this situation, then, we can
neglect the $O(a)$ terms in~Eqs.~\eqref{eq:(3.9e)} and~\eqref{eq:(4.17h)},
because no $O(1/a)$ divergence can arise in correlation functions of a
renormalized operator and elementary fields. Moreover, since $O(\hbar^n)$ is
the lowest non-trivial order of the loop expansion in which the expectation
value becomes non-zero, we can set, for example,
$\langle\mathcal{Z}_T\partial_\mu\Breve{T}_\mu(x)\rangle^{O(\hbar^n)}
=\mathcal{Z}_T^{O(\hbar^n)}
\langle\partial_\mu\Breve{T}_\mu(x)\rangle^{O(\hbar^0)}$. The expectation values
of operators in Eqs.~\eqref{eq:(3.9)} and~\eqref{eq:(4.17)} are thus evaluated
in the tree level approximation and expressions in Eqs.~\eqref{eq:(3.9)}
and~\eqref{eq:(4.17)} can be regarded as the expectation values themselves.

In this way, in the continuum limit, consistency condition~\eqref{eq:(5.39)}
is simplified to
\begin{align}
   &\int d^4x\,\left[
   i\theta\Bar{\xi}\gamma_5X_S(x)^{O(\hbar^n)}
   +\Bar{\xi}rX_S(x)^{O(\hbar^n)}
   -i\theta rX_A(x)^{O(\hbar^n)}
   \right]
\notag\\
   &\qquad{}
   +\int d^4x\,M\left\langle
   \Bar{\psi}^a(x)\left[s^2\psi^a(x)-\gamma_5\xi\Bar{\xi}\gamma_5D\psi^a(x)
   \right]\right\rangle^{O(\hbar^n)}
\notag\\
   &\qquad\qquad{}
   +\int d^4x\,(-i)\Bar{\xi}\gamma_\mu\xi
   \left\langle X_{t_\mu}(x)\right\rangle^{O(\hbar^n)}=0,
\label{eq:(6.3)}
\end{align}
in terms of expressions in~Eqs.~\eqref{eq:(3.9)} and~\eqref{eq:(4.17)} (here we
should not include the $O(a)$ terms, Eqs.~\eqref{eq:(3.9e)}
and~\eqref{eq:(4.17h)}), where
\begin{align}
   rA_\mu(x)&\equiv D_\mu c(x)+\Bar{\xi}\gamma_\mu\psi(x),
\notag\\
   r\psi(x)&\equiv
   -ig\{c(x),\psi(x)\}
   -\frac{1}{2}\sigma_{\mu\nu}\xi F_{\mu\nu}(x)
   +i\theta\gamma_5\psi(x),
\notag\\
   r\Bar{\psi}(x)&\equiv
   -ig\{c(x),\Bar{\psi}(x)\}
   +\frac{1}{2}\Bar{\xi}\sigma_{\mu\nu}F_{\mu\nu}(x)
   +i\theta\Bar{\psi}(x)\gamma_5,
\notag\\
   rc(x)&\equiv
   -igc(x)^2+\Bar{\xi}\gamma_\mu\xi A_\mu(x),
\notag\\
   r\xi&\equiv
   i\theta\gamma_5\xi,\qquad
   r\Bar{\xi}=i\theta\Bar{\xi}\gamma_5,
\notag\\
   rt_\mu&\equiv
   -i\Bar{\xi}\gamma_\mu\xi,
\notag\\
   ru_V(x)&\equiv M,
\notag\\
   rv_A(x)&\equiv-2i\theta M,
\label{eq:(6.4)}
\end{align}
and $r\equiv0$ on other variables. Note
that the $s^2\psi^a$-term in~Eq.~\eqref{eq:(6.3)} is quadratic in~$\xi$, i.e.,
$O(\xi^2)$.

\subsection{$O(\theta^1,\xi^0)$ term}

Relation~\eqref{eq:(6.3)} must hold order by order in~$\theta$ and~$\xi$. Let
us first consider the $O(\theta^1,\xi^0)$ term. This gives
\begin{equation}
   \int d^4x\,\left[rX_A(x)^{O(\hbar^n)}\right]_{O(\theta^0,\xi^0)}=0.
\label{eq:(6.5)}
\end{equation}
When $\theta=\xi=0$, $r$ is nothing but the conventional gauge BRS
transformation in the continuum theory. Thus, Eq.~\eqref{eq:(6.5)} simply
tells that $X_A(x)$ must be (gauge) BRS invariant up to a total
divergence.\footnote{Thus, $X_A(x)$ (and $X_S(x)$ too) is an element of the
local (gauge) BRS cohomology~(see, for example, Ref.~\cite{Barnich:2000zw}). If
one had a complete classification of the cohomology on the basis of lattice
symmetries alone (e.g., the hypercubic symmetry instead of the Lorentz
symmetry), it would greatly simplify the following discussion.} This is
trivially true for~Eqs.~\eqref{eq:(3.9a)}--\eqref{eq:(3.9c)}, but it constrains
the possible structure of the BRS non-invariant piece, Eq.~\eqref{eq:(3.9d)}.

\subsection{$O(\theta^0,\xi^1)$ term}

Similarly, we have from the $O(\theta^0,\xi^1)$ term of~Eq.~\eqref{eq:(6.3)},
\begin{equation}
   \int d^4x\,\left[rX_S(x)^{O(\hbar^n)}\right]_{O(\theta^0,\xi^0)}=0.
\label{eq:(6.6)}
\end{equation}
As an application of this relation, let us ask whether a BRS non-invariant
combination
\begin{equation}
   \gamma_\mu\tr\left[\psi(x)\left\{A_\nu(x),F_{\mu\nu}(x)\right\}\right],
\label{eq:(6.7)}
\end{equation}
can appear in~Eq.~\eqref{eq:(4.17g)} or not. For this operator, we have
\begin{align}
   &\left[r
   \gamma_\mu
   \tr\left[\psi(x)\left\{A_\nu(x),F_{\mu\nu}(x)\right\}\right]
   \right]_{O(\theta^0,\xi^0)}
\notag\\
   &=-\gamma_\mu
   \tr\left[\psi(x)\left\{\partial_\nu c(x),F_{\mu\nu}(x)\right\}\right],
\label{eq:(6.8)}
\end{align}
which is proportional to the totally-symmetric tensor
\begin{equation}
   d^{abc}\equiv\tr\left(T^a\{T^b,T^c\}\right).
\label{eq:(6.9)}
\end{equation}
Then the question is that whether Eq.~\eqref{eq:(6.8)} can be made into a
total-divergence by adding the $r$-transformation of some operator other
than minus Eq.~\eqref{eq:(6.7)}. We see that this is impossible as follows.

Note first that except the term $\partial_\mu c(x)$ in~$rA_\mu(x)$
of~Eq.~\eqref{eq:(6.4)}, $c(x)$ in the $r$-operation appears in the adjoint
action. Since the adjoint action is proportional to the antisymmetric
structure constant, the adjoint action with the ghost~$c(x)$ cannot produce the
totally-symmetric tensor~$d^{abc}$. The only way to form $d^{abc}$ is to use the
$\partial_\mu c(x)$ in~$rA_\mu(x)$. Thus the generic operator to be added
to~Eq.~\eqref{eq:(6.8)} must have the following structure,\footnote{There
exists another possible combination $-r\gamma_\mu E_{\mu\nu\rho\sigma}%
\tr[\partial_\rho\psi(x)\{A_\nu(x),A_\sigma(x)\}]$, but this is reduced to
other combinations in~Eq.~\eqref{eq:(6.10)} up to a total divergence.}
\begin{align}
   &-[r\gamma_\mu\{
   C_{\mu\nu\rho\sigma}
   \tr\left[\psi(x)\left\{A_\nu(x),F_{\rho\sigma}(x)\right\}\right]
\notag\\
   &\qquad\qquad\qquad{}
   +D_{\mu\nu\rho\sigma}
   \tr\left[\psi(x)\left\{A_\nu(x),\partial_\rho A_\sigma(x)\right\}\right]
   \}
   ]_{O(\theta^0,\xi^0)}
\notag\\
   &=\gamma_\mu\{
   C_{\mu\nu\rho\sigma}\tr\left[\psi(x)\left\{\partial_\nu c(x),F_{\rho\sigma}(x)
   \right\}\right]
\notag\\
   &\qquad\qquad{}
   +D_{\mu\nu\rho\sigma}\tr\left[\psi(x)
   \left\{\partial_\nu c(x),\partial_\rho A_\sigma(x)
   \right\}\right]
\notag\\
   &\qquad\qquad\qquad\qquad{}
   +D_{\mu\nu\rho\sigma}\tr\left[\psi(x)
   \left\{A_\nu(x),\partial_\rho\partial_\sigma c(x)
   \right\}\right]\}.
\label{eq:(6.10)}
\end{align}
For the sum of this and Eq.~\eqref{eq:(6.8)} to become a total divergence,
terms proportional to~$\tr[\psi(x)\{\partial_\nu c(x)[A_\rho(x),A_\sigma(x)]\}]$
should cancel out (because it cannot be a total divergence). This requires
$C_{\mu\nu\rho\sigma}=\delta_{\mu\rho}\delta_{\nu\sigma}$. However, then the first
term in~Eq.~\eqref{eq:(6.10)} is just the $r$-transformation of minus
Eq.~\eqref{eq:(6.7)}. This shows that Eq.~\eqref{eq:(6.7)} cannot appear in the
BRS non-invariant piece in $X_S(x)$~\eqref{eq:(4.17g)}; this fact has an
important implication below.

\subsection{$O(\theta^1,\xi^1)$ terms}

At long last, we are now ready to attack one of our main problems. For this,
we take the $O(\theta^1,\xi^1)$ terms of~Eq.~\eqref{eq:(6.3)}:
\begin{align}
   &\int d^4x\,\biggl\{
   i\theta\Bar{\xi}\gamma_5X_S(x)_{O(\theta^0,\xi^0)}^{O(\hbar^n)}
   +\Bar{\xi}\left[
   rX_S(x)^{O(\hbar^n)}
   \right]_{O(\theta^1,\xi^0)}
\notag\\
   &\qquad\qquad\qquad\qquad\qquad{}
   -i\theta\left[
   rX_A(x)^{O(\hbar^n)}
   \right]_{O(\theta^0,\xi^1)}
   \biggr\}=0.
\label{eq:(6.11)}
\end{align}
Then, substituting Eqs.~\eqref{eq:(3.9a)}--\eqref{eq:(3.9c)} and
Eqs.~\eqref{eq:(4.17a)}--\eqref{eq:(4.17e)} in~Eq.~\eqref{eq:(6.11)}, we have
\begin{equation}
   \mathcal{Z}_{\chi}^{O(\hbar^n)}-\frac{1}{2}\mathcal{Z}_P^{O(\hbar^n)}=0,
\label{eq:(6.12)}
\end{equation}
as the coefficient of the combination
\begin{equation}
   -\frac{1}{a}
   \int d^4x\,2i\theta\Bar{\xi}\gamma_5\sigma_{\mu\nu}\tr[\psi(x)F_{\mu\nu}(x)].
\label{eq:(6.13)}
\end{equation}

It is easy to see that the terms containing $c(x)$, $\Bar{c}(x)$ or~$B(x)$,
Eq.~\eqref{eq:(4.17f)}, cannot contribute to~Eq.~\eqref{eq:(6.13)}. We can see
also that BRS non-invariant terms in Eqs.~\eqref{eq:(3.9d)}
and~\eqref{eq:(4.17g)} do not contribute to~Eq.~\eqref{eq:(6.13)} as follows.

First, since the $O(\theta)$ term in~Eq.~\eqref{eq:(6.4)} does not change the
gauge index structure and Eq.~\eqref{eq:(6.13)} is proportional
to~$\delta^{ab}=2\tr(T^aT^b)$, the possible contribution
from~Eq.~\eqref{eq:(4.17g)} to~Eq.~\eqref{eq:(6.13)} comes from
\begin{align}
   &-\frac{1}{a}2
   \Bar{\xi}\biggl[r\biggl\{2F_{\mu\nu\rho\sigma}\sigma_{\mu\nu}
   \tr\left[\psi(x)\partial_\rho A_\sigma(x)\right]
\notag\\
   &\qquad\qquad\qquad{}
   +G_{\mu\nu\rho\sigma}\sigma_{\mu\nu}
   \tr\left[\psi(x)ig\left[A_\rho(x),A_\sigma(x)\right]\right]\biggr\}
   \biggl]_{O(\theta^1,\xi^0)}.
\label{eq:(6.14)}
\end{align}
However, for (the integral of) this to coincide with~Eq.~\eqref{eq:(6.13)} up
to a total divergence, we have to have
$G_{\mu\nu\rho\sigma}=c\delta_{\mu\rho}\delta_{\nu\sigma}$, because the combination
$\tr[\psi(x)[A_\rho(x),A_\sigma(x)]]$ does not contain any derivative. Once this
becomes the case, then, we have to moreover have
$F_{\mu\nu\rho\sigma}=c\delta_{\mu\rho}\delta_{\nu\sigma}$. Then the combination in
the curly brackets of~Eq.~\eqref{eq:(6.14)} is
$c\sigma_{\mu\nu}\tr[\psi(x)F_{\mu\nu}(x)]$, the operator~$(1/a)c\Breve{\chi}(x)$
in~Eq.~\eqref{eq:(4.17c)} that should not be contained
in~Eq.~\eqref{eq:(4.17g)}. On the other hand, concerning Eq.~\eqref{eq:(3.9d)},
the possible combination that could contribute to~Eq.~\eqref{eq:(6.13)} is
\begin{align}
   -i\theta\left[r\left\{
   \frac{1}{a}H_{\mu\rho\sigma}\tr\left[A_\mu(x)F_{\rho\sigma}(x)\right]
   +\frac{1}{a}I_{\mu\rho\sigma}2\tr\left[A_\mu(x)\partial_\rho A_\sigma(x)\right]
   \right\}
   \right]_{O(\theta^0,\xi^1)}.
\label{eq:(6.15)}
\end{align}
However, this cannot contribute to~Eq.~\eqref{eq:(6.13)}, because
$rA_\mu(x)$~\eqref{eq:(6.4)} produces $\Bar{\xi}\gamma_\mu\psi(x)$, but this
combination is linearly independent
of~$\Bar{\xi}\gamma_5\sigma_{\mu\nu}\psi(x)$ (the coefficient
of~Eq.~\eqref{eq:(6.13)}).\footnote{This is because the former changes the sign
under $\xi\to\gamma_5\xi$ and~$\psi(x)\to\gamma_5\psi(x)$, but the latter does
not.} By this way, we see that Eqs.~\eqref{eq:(3.9d)} and~\eqref{eq:(4.17g)} do
not contribute to combination~\eqref{eq:(6.13)}.

Besides these, in the present case, we have to take into consideration also the
possible modification of Eqs.~\eqref{eq:(3.9)} and~\eqref{eq:(4.17)} by the
presence of newly-introduced external variables. From Eqs.~\eqref{eq:(6.4)},
we see that if there exist mixings such that
\begin{align}
   X_S(x)&\sim-\frac{1}{a}\frac{u_V(x)}{M}
   2i\theta\gamma_5\sigma_{\mu\nu}
   \tr\left[\psi(x)F_{\mu\nu}(x)\right],
\notag\\
   X_S(x)&\sim\frac{1}{a}\frac{v_A(x)}{M}
   \gamma_5\sigma_{\mu\nu}
   \tr\left[\psi(x)F_{\mu\nu}(x)\right],
\notag\\
   X_A(x)&\sim\frac{1}{a}\frac{u_V(x)}{M}
   2\Bar{\xi}\gamma_5\sigma_{\mu\nu}
   \tr\left[\psi(x)F_{\mu\nu}(x)\right],
\label{eq:(6.16)}
\end{align}
then these produce combination~\eqref{eq:(6.13)} under~$r$. These possibilities
are excluded, however, because in our new total action~$S_{\text{tot}}$, $M$,
$v_V(x)$ and~$v_A(x)$ appear always in the particular combination,
$M+v_V(x)+v_A(x)\gamma_5$. In this way, we establishes
relation~\eqref{eq:(6.12)}.

In Eq.~\eqref{eq:(6.12)}, the integer~$n\geq1$ was the lowest-order of the loop
expansion in which $\mathcal{Z}_{\chi}^{O(\hbar^n)}\neq0$
and~$\mathcal{Z}_P^{O(\hbar^n)}\neq0$. However, once Eq.~\eqref{eq:(6.12)} is
fulfilled, then the consistency condition~\eqref{eq:(6.11)} applies to the next
leading order~$O(\hbar^{n+1})$. Thus, repeating the above argument, we have
\begin{equation}
   \mathcal{Z}_{\chi}=\frac{1}{2}\mathcal{Z}_P,
\label{eq:(6.17)}
\end{equation}
to all orders of the loop expansion. This proves one of our assertions,
Eq.~\eqref{eq:(4.22)}.

\subsection{$O(\theta^0,\xi^2)$ terms}

Finally, we consider the $O(\theta^0,\xi^2)$ terms in~Eq.~\eqref{eq:(6.3)}.
This yields,
\begin{align}
   &\int d^4x\,
   \Bar{\xi}\left[rX_S(x)^{O(\hbar^n)}
   \right]_{O(\theta^0,\xi^1)}
\notag\\
   &\qquad{}
   +\int d^4x\,M\left\langle
   \Bar{\psi}^a(x)\left[s^2\psi^a(x)-\gamma_5\xi\Bar{\xi}\gamma_5D\psi^a(x)
   \right]\right\rangle_{O(\theta^0,\xi^2)}^{O(\hbar^n)}
\notag\\
   &\qquad\qquad{}
   +\int d^4x\,(-i)\Bar{\xi}\gamma_\mu\xi
   \left\langle X_{t_\mu}(x)\right\rangle^{O(\hbar^n)}_{O(\theta^0,\xi^0)}=0.
\label{eq:(6.18)}
\end{align}
Then, if we substitute Eqs.~\eqref{eq:(4.17a)}--\eqref{eq:(4.17e)} in the
first term, we have
\begin{equation}
   \mathcal{Z}_{3F}^{O(\hbar^n)}=0,
\label{eq:(6.19)}
\end{equation}
as the coefficient of the combination\footnote{We have noted that the
coefficient~$\Bar{\xi}\sigma_{\mu\nu}\xi$ is linearly independent
of~$\Bar{\xi}\gamma_\mu\xi$ (the coefficient of the last term
of~Eq.~\eqref{eq:(6.18)}) because under~$\xi\to\gamma_5\xi$, the former does
not change the sign but the latter does.}
\begin{equation}
   \frac{1}{4}\int d^4x\,\left\{
   \Bar{\xi}\sigma_{\mu\nu}\xi
   \tr\left[F_{\mu\nu}(x)\Bar{\psi}(x)\psi(x)\right]
   -\Bar{\xi}\gamma_5\sigma_{\mu\nu}\xi
   \tr\left[F_{\mu\nu}(x)\Bar{\psi}(x)\gamma_5\psi(x)\right]
   \right\}.
\label{eq:(6.20)}
\end{equation}
In fact, it is easy to see without any calculation that among
Eqs.~\eqref{eq:(4.17a)}--\eqref{eq:(4.17e)} only the three-fermion
operator~\eqref{eq:(4.17d)} contributes to~Eq.~\eqref{eq:(6.20)}:
Eq.~\eqref{eq:(4.17d)} is proportional to the totally-symmetric
tensor~\eqref{eq:(6.9)}, while others are proportional
to~$\delta^{ab}=2\tr(T^aT^b)$.

Again, it is easy to see that the terms containing $c(x)$, $\Bar{c}(x)$
or~$B(x)$, Eq.~\eqref{eq:(4.17f)}, do not contribute to~Eq.~\eqref{eq:(6.20)}.

Concerning the possible contribution of the BRS non-invariant
terms~\eqref{eq:(4.17g)} to~Eq.~\eqref{eq:(6.20)}, the unique possibility is
\begin{equation}
   \Bar{\xi}
   \left[
   r\frac{1}{2}J_{\mu\nu\rho\sigma}
   \gamma_\mu\tr\left[\psi(x)\left\{A_\nu(x),F_{\rho\sigma}(x)\right\}\right]
   \right]_{O(\theta^0,\xi^1)}.
\label{eq:(6.21)}
\end{equation}
One sees that when $J_{\mu\nu\rho\sigma}=d\delta_{\mu\rho}\delta_{\nu\sigma}$, (the
integral of) this is proportional to~Eq.~\eqref{eq:(6.20)} by using the Fierz
theorem~\eqref{eq:(A.1)}. However, we have shown in~Eq.~\eqref{eq:(6.7)} that
Eq.~\eqref{eq:(6.21)}
with~$J_{\mu\nu\rho\sigma}=d\delta_{\mu\rho}\delta_{\nu\sigma}$ cannot appear in
the mixing of~$X_S(x)$~\eqref{eq:(4.17g)}. Thus, BRS non-invariant
operators~\eqref{eq:(4.17g)} do not contribute to~Eq.~\eqref{eq:(6.20)}.

For the modification of Eqs.~\eqref{eq:(4.17)} in the presence of new external
variables, what could contribute to~Eq.~\eqref{eq:(6.20)} is the combination
\begin{align}
   &X_S(x)
\notag\\
   &\sim\frac{u_V(x)}{M}
   \frac{1}{4}\left\{
   \sigma_{\mu\nu}\xi
   \tr\left[F_{\mu\nu}(x)\Bar{\psi}(x)\psi(x)\right]
   -\gamma_5\sigma_{\mu\nu}\xi
   \tr\left[F_{\mu\nu}(x)\Bar{\psi}(x)\gamma_5\psi(x)\right]
   \right\}.
\label{eq:(6.22)}
\end{align}
This possibility is again excluded, because $u_V(x)$ should appear only in the
combination~$M+u_V(x)$. In this way, we establish Eq.~\eqref{eq:(6.19)}, and
repeating the argument for higher~$n$, we have
\begin{equation}
   \mathcal{Z}_{3F}=0.
\label{eq:(6.23)}
\end{equation}
That is, we establish Eq.~\eqref{eq:(4.23)}.

\section{Conclusion}
\label{sec:7}

In this paper, in the context of the lattice regularization of 4D
$\mathcal{N}=1$ SYM, we formulated a generalized BRS transformation that treats
the gauge, SUSY, translation and $U(1)_A$ transformations in a unified way. On
the basis of this (almost-nilpotent) transformation on lattice variables, we
obtained a generalized WZ consistency condition for the symmetry breaking
effects in the lattice formulation. Utilizing this powerful machinery, we then
proved that $\mathcal{Z}_\chi=(1/2)\mathcal{Z}_P$ which implies the coincidence
of the chiral symmetric limit~\eqref{eq:(3.13)} and the supersymmetric
limit~\eqref{eq:(4.21)}, and that $\mathcal{Z}_{3F}=0$ which implies that there
is no exotic breaking of the SUSY WT identity by the three-fermion
operator~\eqref{eq:(4.19)}. Our these results provide a solid theoretical basis
for lattice formulations of 4D $\mathcal{N}=1$ SYM. It is interesting to
investigate further consequences of the consistency condition~\eqref{eq:(5.39)}
or~\eqref{eq:(6.3)}.

In the continuum theory, the generalized BRS symmetry has been formulated
also for 4D $\mathcal{N}=4$ SYM~\cite{White:1992wu} and for 4D $\mathcal{N}=2$
SYM~\cite{Maggiore:1994xw,Maggiore:1994dw}; in the former, such a framework
is crucially important because no off-shell multiplet is known. Adopting this
framework to the lattice formulation might be useful to systematically classify
the necessary fine-tuning in such formulations.\footnote{This problem has been
studied for 4D $\mathcal{N}=4$ SYM in a recent paper~\cite{Catterall:2011pd}.}

\section*{Acknowledgements}

I am indebted to Yusuke Taniguchi for helpful discussions at various stages of
the present work and to Michael G.~Endres for a careful reading of the
manuscript and useful comments. This work was initiated during the CERN Theory
Institute, ``Future directions in lattice gauge theory---LGT10''. I would like
to thank the workshop organizers, especially, Poul H.~Damgaard and Martin
L\"uscher, for their kind hospitality. My thanks also goes to Ting-Wai Chiu for
the hospitality extended to me at the Center for Theoretical Sciences, Taipei,
Taiwan, R.O.C., where the latter part of this work was carried out. This work
is supported in part by a Grant-in-Aid for Scientific Research, 22340069
and~23540330.

\section*{Note added in the proof}
The operator mixing coefficients $\mathcal{Z}_P$ and~$\mathcal{Z}_\chi$ are
dimensionless combinations of the bare gauge coupling constant~$g$ and the bare
gluino mass~$M$,\footnote{These power-divergence subtraction coefficients are
independent of the renormalization point~$\mu$~\cite{Testa:1998ez}.}
\begin{equation}
   \mathcal{Z}_P=\mathcal{Z}_P(g,aM),\qquad
   \mathcal{Z}_\chi=\mathcal{Z}_\chi(g,aM).
\label{eq:(7.1)}
\end{equation}
In the perturbation theory in the present paper, the mass~$M$ (not $aM$) is
treated as a fixed parameter and the combination $aM$ is hence regarded as a
higher-order quantity in the lattice spacing. What we have proven
in~Eq.~\eqref{eq:(6.17)} is thus the equality
\begin{equation}
   \mathcal{Z}_\chi(g,0)=\frac{1}{2}\mathcal{Z}_P(g,0),
\label{eq:(7.2)}
\end{equation}
to all orders in the power series of~$g$.

On the other hand, in actual numerical simulations with a small but fixed
lattice spacing~$a$, the chiral limit would be specified by the tuning,
$M\to M_{\text{cr}}^{\text{chiral}}(g)$, where $M_{\text{cr}}^{\text{chiral}}(g)$ is
the solution of~\cite{Bochicchio:1985xa}
\begin{equation}
   aM_{\text{cr}}^{\text{chiral}}(g)
   -\frac{1}{2}\mathcal{Z}_P(g,aM_{\text{cr}}^{\text{chiral}}(g))=0.
\label{eq:(7.3)}
\end{equation}
Similarly, the supersymmetric limit would be specified by,
$M\to M_{\text{cr}}^{\text{SUSY}}(g)$, where~\cite{Donini:1997hh}
\begin{equation}
   aM_{\text{cr}}^{\text{SUSY}}(g)
   -\mathcal{Z}_\chi(g,aM_{\text{cr}}^{\text{SUSY}}(g))=0.
\label{eq:(7.4)}
\end{equation}
Then the question is whether from~Eq.~\eqref{eq:(7.2)} one can draw any
conclusion concerning the relation between $M_{\text{cr}}^{\text{chiral}}(g)$
and~$M_{\text{cr}}^{\text{SUSY}}(g)$ which are defined by Eqs.~\eqref{eq:(7.3)}
and~\eqref{eq:(7.4)}, respectively. The answer is affirmative as follows: The
point is that the mixing coefficients $\mathcal{Z}_P$ and~$\mathcal{Z}_\chi$
are of~$O(g^2)$, thus for a fixed $aM$, $\mathcal{Z}_P(g,aM)$
and~$\mathcal{Z}_\chi(g,aM)$ become arbitrarily small as $g\to0$. This shows
that, irrespective of how $\mathcal{Z}_P$ and~$\mathcal{Z}_\chi$ depend
on~$aM$, we have
\begin{equation}
   \lim_{g\to0}aM_{\text{cr}}^{\text{chiral}}(g)
  =\lim_{g\to0}aM_{\text{cr}}^{\text{chiral}}(g)=0. 
\label{eq:(7.5)}
\end{equation}
In particular, $aM_{\text{cr}}^{\text{chiral}}(g)$
and~$aM_{\text{cr}}^{\text{SUSY}}(g)$ vanish in the quantum continuum limit in
which $g\to0$. This allows us to approximate Eqs.~\eqref{eq:(7.3)}
and~\eqref{eq:(7.4)} by
\begin{align}
   aM_{\text{cr}}^{\text{chiral}}(g)
   -\frac{1}{2}\mathcal{Z}_P(g,0)&=0,
\notag\\
   aM_{\text{cr}}^{\text{SUSY}}(g)
   -\mathcal{Z}_\chi(g,0)&=0,
\end{align}
as we approach to the continuum limit. Then Eq.~\eqref{eq:(7.2)} implies
that $M_{\text{cr}}^{\text{chiral}}(g)$ and~$M_{\text{cr}}^{\text{SUSY}}(g)$
coincide in the continuum limit.

\appendix

\section{Three-fermion spinorial operators}
\label{sec:A}

In this appendix, we explore the relations between various gauge-invariant
three-fermion operators and show that the
operator~$\mathcal{O}_S$~\eqref{eq:(4.19)} generally does not vanish.

For any $4\times4$ matrices $\Lambda^1$ and~$\Lambda^2$, we have the Fierz
theorem
\begin{align}
   \Lambda^1\psi^a\left(\Bar{\psi}^b\Lambda^2\psi^c\right)
   &=\mp\frac{1}{4}\sum_{\lambda_A}\lambda_A\psi^c
   \left(\Bar{\psi}^b\Lambda^2\lambda_A\Lambda^1\psi^a\right)
\notag\\
   &=\frac{1}{4}\sum_{\lambda_A}\lambda_A\psi^c
   \left(\psi^{Ta}\Lambda^{1T}\lambda_A^T\Lambda^{2T}C^{-1}
   \psi^b\right),
\label{eq:(A.1)}
\end{align}
where the upper sign holds when both $\psi^a$ and~$\psi^b$ are Grassmann-odd
and the lower holds otherwise; $\lambda_A$ is the complete basis for~$4\times4$
matrices:
\begin{equation}
   \lambda_A\equiv\left\{1,\gamma_5,\gamma_\alpha,i\gamma_5\gamma_\alpha,
   i\sigma_{\alpha\beta}\right\}.
\label{eq:(A.2)}
\end{equation}
In the last line of~Eq.~\eqref{eq:(A.1)}, we have used the
constraint~$\Bar{\psi}^b=\psi^{Tb}(-C^{-1})$ and~$C^T=-C$. Since
\begin{equation}
   \lambda_A^T\equiv c_AC^{-1}\lambda_AC,
\label{eq:(A.3)}
\end{equation}
where
\begin{equation}
   c_A=\left\{1,1,-1,1,-1\right\},
\label{eq:(A.4)}
\end{equation}
when $\Lambda^1=\Lambda^2=\Lambda\in\lambda_A$, we have
\begin{equation}
   \Lambda\psi^a\left(\Bar{\psi}^b\Lambda\psi^c\right)
   =-\frac{1}{4}\sum_Ac_A\lambda_A\psi^c
   \left(\Bar{\psi}^a\Lambda\lambda_A\Lambda
   \psi^b\right).
\label{eq:(A.5)}
\end{equation}
Multiplying by $\tr\left(T^aT^bT^c\right)$ and summing over $a$, $b$,
and~$c$, we have
\begin{equation}
   \Lambda\tr\left[\psi\left(\Bar{\psi}\Lambda\psi\right)\right]
   =-\frac{1}{4}\sum_Ac_A\lambda_A\tr\left[\psi
   \left(\Bar{\psi}\Lambda\lambda_A\Lambda
   \psi\right)\right].
\label{eq:(A.6)}
\end{equation}
Applying this relation to
\begin{align}
   \mathcal{O}_S
   &\equiv\tr\left[\psi\left(\Bar{\psi}\psi\right)\right],
\notag\\
   \mathcal{O}_P
   &\equiv\gamma_5\tr\left[\psi\left(\Bar{\psi}\gamma_5\psi\right)\right],
\notag\\
   \mathcal{O}_V
   &\equiv\sum_\mu\gamma_\mu
   \tr\left[\psi\left(\Bar{\psi}\gamma_\mu\psi\right)\right],
\notag\\
   \mathcal{O}_A
   &\equiv\sum_\mu i\gamma_5\gamma_\mu
   \tr\left[\psi\left(\Bar{\psi}i\gamma_5\gamma_\mu\psi\right)\right],
\notag\\
   \mathcal{O}_T
   &\equiv\sum_{\mu<\nu}i\sigma_{\mu\nu}
   \tr\left[\psi\left(\Bar{\psi}i\sigma_{\mu\nu}\psi\right)\right],
\label{eq:(A.7)}
\end{align}
and using
\begin{align}
   1\lambda_A1
   &\equiv s_A\lambda_A,&&s_A=\left\{1,1,1,1,1\right\},
\notag\\
   \gamma_5\lambda_A\gamma_5
   &\equiv p_A\lambda_A,&&p_A=\left\{1,1,-1,-1,1\right\},
\notag\\
   \sum_\mu\gamma_\mu\lambda_A\gamma_\mu
   &\equiv v_A\lambda_A,&&v_A=\left\{4,-4,-2,2,0\right\},
\notag\\
   \sum_\mu i\gamma_5\gamma_\mu\lambda_Ai\gamma_5\gamma_\mu
   &\equiv a_A\lambda_A,&&a_A=\left\{4,-4,2,-2,0\right\},
\notag\\
   \sum_{\mu<\nu}i\sigma_{\mu\nu}\lambda_Ai\sigma_{\mu\nu}
   &\equiv t_A\lambda_A,&&t_A=\left\{6,6,0,0,-2\right\},
\end{align}
we have
\begin{align}
   &\mathcal{O}_S=-\frac{1}{4}\sum_Ac_As_A\lambda_A
   \tr\left[\psi\left(\Bar{\psi}\lambda_A\psi\right)\right],
\notag\\
   &\mathcal{O}_P=-\frac{1}{4}\sum_Ac_Ap_A\lambda_A
   \tr\left[\psi\left(\Bar{\psi}\lambda_A\psi\right)\right],
\notag\\
   &\mathcal{O}_V=-\frac{1}{4}\sum_Ac_Av_A\lambda_A
   \tr\left[\psi\left(\Bar{\psi}\lambda_A\psi\right)\right],
\notag\\
   &\mathcal{O}_A=-\frac{1}{4}\sum_Ac_Aa_A\lambda_A
   \tr\left[\psi\left(\Bar{\psi}\lambda_A\psi\right)\right],
\notag\\
   &\mathcal{O}_T=-\frac{1}{4}\sum_Ac_At_A\lambda_A
   \tr\left[\psi\left(\Bar{\psi}\lambda_A\psi\right)\right],
\end{align}
or, equivalently,
\begin{equation}
   -4
   \begin{pmatrix}
   \mathcal{O}_S\\\mathcal{O}_P\\\mathcal{O}_V\\\mathcal{O}_A\\\mathcal{O}_T
   \end{pmatrix}
   =
   \begin{pmatrix}
   1&1&-1&1&-1\\
   1&1&1&-1&-1\\
   4&-4&2&2&0\\
   4&-4&-2&-2&0\\
   6&6&0&0&2
   \end{pmatrix}
   \begin{pmatrix}
   \mathcal{O}_S\\\mathcal{O}_P\\\mathcal{O}_V\\\mathcal{O}_A\\\mathcal{O}_T
   \end{pmatrix}.
\end{equation}
By solving this (singular) simultaneous equation, we find
\begin{equation}
   \mathcal{O}_V=\mathcal{O}_T=0,\qquad
   \mathcal{O}_P=-\mathcal{O}_S,\qquad
   \mathcal{O}_A=-4\mathcal{O}_S,
\end{equation}
and $\mathcal{O}_S$ is undetermined.

With a help of a Mathematica package~\cite{grassmann.m}, we verified that the
combination $\mathcal{O}_S$ is in fact non-zero for the gauge group~$SU(3)$, by
explicitly expressing it in terms of Grassmann-odd variables~$\psi_\alpha^a$.
(When the gauge group is $SU(2)$, one immediately sees that $\mathcal{O}_S=0$
and hence $\mathcal{O}_P=\mathcal{O}_A=0$.) Thus the combinations,
$\mathcal{O}_S$, $\mathcal{O}_P$ and~$\mathcal{O}_A$, can generally be
non-vanishing.

\section{Triviality of dimension $9/2$ $s_0$-invariant operators that
contain $c$, $\Bar{c}$ or~$B$}
\label{sec:B}

In this appendix, we show the following\footnote{A corresponding statement
(without a proof) can be found in~Ref.~\cite{Curci:1986sm} (just
below~Eq.~(17)).}
\begin{lem}
Suppose $\mathcal{O}(x)$ is an $s_0$-invariant operator with zero ghost-number
that contains a gauge ghost~$c$, an anti-ghost~$\Bar{c}$ or an auxiliary
field~$B$. If its mass-dimension is~$9/2$ or less and if it behaves in the same
way as~$\chi(x)$~\eqref{eq:(4.10)} under lattice discrete symmetries,
$\mathcal{O}(x)$ is $s_0$-exact.
\end{lem}
\begin{proof}
To comply with the behavior under the hypercubic transformation,
$\mathcal{O}(x)$ must contain at least one~$\psi(x)$ whose mass-dimension
is~$3/2$. From this and the fact that $\mathcal{O}(x)$ is of zero
ghost-number, it immediately follows that $\mathcal{O}(x)$ is linear
in~$\Bar{c}(x)$ or in~$B(x)$. Thus, the most general form of~$\mathcal{O}(x)$
is given by
\begin{equation}
   \mathcal{O}(x)
   =\Bar{c}^a(x)\Delta_1^a(x)-B^a(x)\Delta_0^a(x)
   +\partial_\mu\Bar{c}^a(x)\Delta_{1\mu}^a(x)
   -\partial_\mu B^a(x)\Delta_{0\mu}^a(x),
\end{equation}
where $\Delta$'s are independent of~$\Bar{c}$. The $s_0$-invariance
$s_0\mathcal{O}(x)=0$ of course implies
$\left.s_0\mathcal{O}(x)\right|_{\Bar{c}=0}=0$ and, from this, we have
$\Delta_1^a(x)=s_0\Delta_0^a(x)$ and $\Delta_{1\mu}^a(x)=s_0\Delta_{0\mu}^a(x)$.
Hence,
\begin{equation}
   \mathcal{O}(x)
   =-s_0\left[
   \Bar{c}^a(x)\Delta_0^a(x)
   +\partial_\mu\Bar{c}^a(x)\Delta_{0\mu}^a(x)\right].
\end{equation}
\end{proof}

\section{Useful identities}
\label{sec:C}

For a Grassmann-even spinor that obeys constraint~\eqref{eq:(5.3)}, the
following relations hold:
\begin{equation}
   \Bar{\xi}\xi=\Bar{\xi}\gamma_5\xi=\Bar{\xi}\gamma_5\gamma_\mu\xi=0,
\label{eq:(C.1)}
\end{equation}
and
\begin{align}
   \Bar{\xi}\gamma_5\sigma_{\mu\nu}\xi
   &=-\epsilon_{\mu\nu\rho\sigma}\Bar{\xi}\sigma_{\rho\sigma}\xi,
\notag\\
   \Bar{\xi}\gamma_\mu\gamma_\nu\gamma_\rho\xi
   &=\Bar{\xi}\left(
   \delta_{\nu\rho}\gamma_\mu
   -\delta_{\mu\rho}\gamma_\nu
   +\delta_{\mu\nu}\gamma_\rho
   \right)\xi,
\notag\\
   \Bar{\xi}\gamma_5\gamma_\mu\gamma_\nu\gamma_\rho\xi
   &=2\epsilon_{\mu\nu\rho\sigma}\Bar{\xi}\gamma_\sigma\xi,
\notag\\
   \Bar{\xi}\gamma_\mu\gamma_\nu\gamma_\rho\gamma_\sigma\xi
   &=\Bar{\xi}\left(
   \delta_{\mu\nu}\sigma_{\rho\sigma}
   -\delta_{\mu\rho}\sigma_{\nu\sigma}
   +\delta_{\mu\sigma}\sigma_{\nu\rho}
   +\delta_{\nu\rho}\sigma_{\mu\sigma}
   -\delta_{\nu\sigma}\sigma_{\mu\rho}
   +\delta_{\rho\sigma}\sigma_{\mu\nu}
   \right)\xi.
\end{align}

On the other hand, for a Grassmann-odd spinor that obeys
constraint~\eqref{eq:(2.7)}, we have
\begin{equation}
   S^{ab}\Bar{\psi}^a\gamma_\mu\psi^b
   =S^{ab}\Bar{\psi}^a\sigma_{\mu\nu}\psi^b
   =S^{ab}\Bar{\psi}^a\gamma_5\sigma_{\mu\nu}\psi^b=0,
\end{equation}
for a symmetric coefficient $S^{ba}=S^{ab}$, and
\begin{equation}
   A^{ab}\Bar{\psi}^a\psi^b
   =A^{ab}\Bar{\psi}^a\gamma_5\psi^b
   =A^{ab}\Bar{\psi}^a\gamma_5\gamma_\mu\psi^b=0,
\end{equation}
for an antisymmetric coefficient $A^{ba}=-A^{ab}$.




\bibliographystyle{elsarticle-num}
\bibliography{<your-bib-database>}

\begin{thebibliography}{00}


\bibitem{Curci:1986sm}
  G.~Curci and G.~Veneziano,
  Nucl.\ Phys.\ B {\bf 292} (1987) 555.

\bibitem{Kaplan:1983sk}
  D.~B.~Kaplan,
  Phys.\ Lett.\ B {\bf 136} (1984) 162.

\bibitem{Montvay:1996pz}
  I.~Montvay,
  Nucl.\ Phys.\ Proc.\ Suppl.\  {\bf 53} (1997) 853
  [hep-lat/9607035].

\bibitem{Montvay:1997ak}
  I.~Montvay,
  Nucl.\ Phys.\ Proc.\ Suppl.\  {\bf 63} (1998) 108
  [hep-lat/9709080].

\bibitem{Koutsoumbas:1997de}
  G.~Koutsoumbas, I.~Montvay, A.~Pap, K.~Spanderen, D.~Talkenberger and J.~Westphalen,
  Nucl.\ Phys.\ Proc.\ Suppl.\  {\bf 63} (1998) 727
  [hep-lat/9709091].

\bibitem{Kirchner:1998nk}
  R.~Kirchner {\it et al.}  [DESY-M\"unster Collaboration],
  Nucl.\ Phys.\ Proc.\ Suppl.\  {\bf 73} (1999) 828
  [hep-lat/9808024].

\bibitem{Kirchner:1998mp}
  R.~Kirchner {\it et al.}  [DESY-M\"unster Collaboration],
  Phys.\ Lett.\ B {\bf 446} (1999) 209
  [hep-lat/9810062].

\bibitem{Campos:1999du}
  I.~Campos {\it et al.}  [DESY-M\"unster Collaboration],
  Eur.\ Phys.\ J.\ C {\bf 11} (1999) 507
  [hep-lat/9903014].

\bibitem{Feo:1999hw}
  A.~Feo {\it et al.}  [DESY-M\"unster Collaboration],
  Nucl.\ Phys.\ Proc.\ Suppl.\  {\bf 83} (2000) 661
  [hep-lat/9909070].

\bibitem{Feo:1999hx}
  A.~Feo {\it et al.}  [DESY-M\"unster Collaboration],
  Nucl.\ Phys.\ Proc.\ Suppl.\  {\bf 83} (2000) 670
  [hep-lat/9909071].

\bibitem{Farchioni:2000mp}
  F.~Farchioni, A.~Feo, T.~Galla, C.~Gebert, R.~Kirchner, I.~Montvay, G.~M\"unster and A.~Vladikas,
  Nucl.\ Phys.\ Proc.\ Suppl.\  {\bf 94} (2001) 787
  [hep-lat/0010053].

\bibitem{Farchioni:2000kb}
  F.~Farchioni, A.~Feo, T.~Galla, C.~Gebert, R.~Kirchner, I.~Montvay and G.~M\"unster,
  Nucl.\ Phys.\ Proc.\ Suppl.\  {\bf 94} (2001) 791
  [hep-lat/0011030].

\bibitem{Farchioni:2001yn}
  F.~Farchioni, A.~Feo, T.~Galla, C.~Gebert, R.~Kirchner, I.~Montvay, G.~M\"unster and A.~Vladikas,
  Nucl.\ Phys.\ Proc.\ Suppl.\  {\bf 106} (2002) 938
  [hep-lat/0110110].

\bibitem{Farchioni:2001yr}
  F.~Farchioni, A.~Feo, T.~Galla, C.~Gebert, R.~Kirchner, I.~Montvay, G.~M\"unster and R.~Peetz {\it et al.},
  Nucl.\ Phys.\ Proc.\ Suppl.\  {\bf 106} (2002) 941
  [hep-lat/0110113].

\bibitem{Farchioni:2001wx}
  F.~Farchioni {\it et al.}  [DESY-M\"unster-Roma Collaboration],
  Eur.\ Phys.\ J.\ C {\bf 23} (2002) 719
  [hep-lat/0111008].

\bibitem{Peetz:2002sr}
  R.~Peetz, F.~Farchioni, C.~Gebert and G.~M\"unster,
  Nucl.\ Phys.\ Proc.\ Suppl.\  {\bf 119} (2003) 912
  [hep-lat/0209065].

\bibitem{Farchioni:2004fy}
  F.~Farchioni and R.~Peetz,
  Eur.\ Phys.\ J.\ C {\bf 39} (2005) 87
  [hep-lat/0407036].

\bibitem{Demmouche:2008ms}
  K.~Demmouche, F.~Farchioni, A.~Ferling, G.~M\"unster, J.~Wuilloud, I.~Montvay and E.~E.~Scholz,
  PoS LATTICE {\bf 2008} (2008) 061
  [arXiv:0810.0144 [hep-lat]].

\bibitem{Demmouche:2008aq}
  K.~Demmouche, F.~Farchioni, A.~Ferling, G.~M\"unster, J.~Wuilloud, I.~Montvay and E.~E.~Scholz,
  PoS CONFINEMENT {\bf 8} (2008) 136
  [arXiv:0811.1964 [hep-lat]].

\bibitem{Demmouche:2009ki}
  K.~Demmouche, F.~Farchioni, A.~Ferling, I.~Montvay, G.~M\"unster, E.~E.~Scholz and J.~Wuilloud,
  PoS LAT {\bf 2009} (2009) 268
  [arXiv:0911.0595 [hep-lat]].

\bibitem{Demmouche:2010sf}
  K.~Demmouche, F.~Farchioni, A.~Ferling, I.~Montvay, G.~M\"unster, E.~E.~Scholz and J.~Wuilloud,
  Eur.\ Phys.\ J.\ C {\bf 69} (2010) 147
  [arXiv:1003.2073 [hep-lat]].

\bibitem{Bergner:2011wf}
  G.~Bergner, I.~Montvay, G.~M\"unster, D.~Sandbrink and U.~D.~\"Ozugurel,
  arXiv:1111.3012 [hep-lat].

\bibitem{Wilson:1975hf}
  K.~G.~Wilson,
  Subnucl.\ Ser.\  {\bf 13} (1977) 13.

\bibitem{Montvay:2001aj}
  I.~Montvay,
  Int.\ J.\ Mod.\ Phys.\ A {\bf 17} (2002) 2377
  [hep-lat/0112007].

\bibitem{Fleming:2000fa}
  G.~T.~Fleming, J.~B.~Kogut and P.~M.~Vranas,
  Phys.\ Rev.\ D {\bf 64} (2001) 034510
  [hep-lat/0008009].

\bibitem{Giedt:2008xm}
  J.~Giedt, R.~Brower, S.~Catterall, G.~T.~Fleming and P.~Vranas,
  Phys.\ Rev.\ D {\bf 79} (2009) 025015
  [arXiv:0810.5746 [hep-lat]].

\bibitem{Endres:2009yp}
  M.~G.~Endres,
  Phys.\ Rev.\ D {\bf 79} (2009) 094503
  [arXiv:0902.4267 [hep-lat]].

\bibitem{Endres:2009pu}
  M.~G.~Endres,
  PoS LAT {\bf 2009} (2009) 053
  [arXiv:0912.0207 [hep-lat]].

\bibitem{Kaplan:1992bt}
  D.~B.~Kaplan,
  Phys.\ Lett.\ B {\bf 288} (1992) 342
  [hep-lat/9206013].

\bibitem{Shamir:1993zy}
  Y.~Shamir,
  Nucl.\ Phys.\ B {\bf 406} (1993) 90
  [hep-lat/9303005].

\bibitem{Kim:2011fw}
  S.~W.~Kim {\it et al.}  [The JLQCD Collaboration],
  arXiv:1111.2180 [hep-lat].

\bibitem{Neuberger:1997fp}
  H.~Neuberger,
  Phys.\ Lett.\ B {\bf 417} (1998) 141
  [hep-lat/9707022].

\bibitem{Neuberger:1998wv}
  H.~Neuberger,
  Phys.\ Lett.\ B {\bf 427} (1998) 353
  [hep-lat/9801031].

\bibitem{Nishimura:1997vg}
  J.~Nishimura,
  Phys.\ Lett.\ B {\bf 406} (1997) 215
  [hep-lat/9701013].

\bibitem{Maru:1997kh}
  N.~Maru and J.~Nishimura,
  Int.\ J.\ Mod.\ Phys.\ A {\bf 13} (1998) 2841
  [hep-th/9705152].

\bibitem{Donini:1997hh}
  A.~Donini, M.~Guagnelli, P.~Hernandez and A.~Vladikas,
  Nucl.\ Phys.\ B {\bf 523} (1998) 529
  [hep-lat/9710065].

\bibitem{Taniguchi:1999fc}
  Y.~Taniguchi,
  Phys.\ Rev.\ D {\bf 63} (2000) 014502
  [hep-lat/9906026].

\bibitem{Kaplan:1999jn}
  D.~B.~Kaplan and M.~Schmaltz,
  Chin.\ J.\ Phys.\  {\bf 38} (2000) 543
  [hep-lat/0002030].


\bibitem{Mehta:2011ud}
  D.~Mehta, S.~Catterall, R.~Galvez and A.~Joseph,
  arXiv:1112.5413 [hep-lat].

\bibitem{Karsten:1980wd}
  L.~H.~Karsten and J.~Smit,
  Nucl.\ Phys.\ B {\bf 183} (1981) 103.

\bibitem{Bochicchio:1985xa}
  M.~Bochicchio, L.~Maiani, G.~Martinelli, G.~C.~Rossi and M.~Testa,
  Nucl.\ Phys.\ B {\bf 262} (1985) 331.

\bibitem{Testa:1998ez}
  M.~Testa,
  JHEP {\bf 9804} (1998) 002
  [hep-th/9803147].

\bibitem{Ginsparg:1981bj}
  P.~H.~Ginsparg and K.~G.~Wilson,
  Phys.\ Rev.\ D {\bf 25} (1982) 2649.

\bibitem{Sugino:2003yb}
  F.~Sugino,
  JHEP {\bf 0401} (2004) 015
  [hep-lat/0311021].

\bibitem{Sugino:2004qd}
  F.~Sugino,
  JHEP {\bf 0403} (2004) 067
  [hep-lat/0401017].

\bibitem{Kadoh:2009rw}
  D.~Kadoh and H.~Suzuki,
  Phys.\ Lett.\ B {\bf 682} (2010) 466
  [arXiv:0908.2274 [hep-lat]].

\bibitem{Kanamori:2008bk}
  I.~Kanamori and H.~Suzuki,
  Nucl.\ Phys.\ B {\bf 811} (2009) 420
  [arXiv:0809.2856 [hep-lat]].

\bibitem{Ferrara:1974pz}
  S.~Ferrara and B.~Zumino,
  Nucl.\ Phys.\ B {\bf 87} (1975) 207.

\bibitem{White:1992ai}
  P.~L.~White,
  Class.\ Quant.\ Grav.\  {\bf 9} (1992) 1663.

\bibitem{Maggiore:1995gr}
  N.~Maggiore, O.~Piguet and S.~Wolf,
  Nucl.\ Phys.\ B {\bf 458} (1996) 403
   [Erratum-ibid.\ B {\bf 469} (1996) 513]
  [hep-th/9507045].

\bibitem{Maggiore:1996gg}
  N.~Maggiore, O.~Piguet and S.~Wolf,
  Nucl.\ Phys.\ B {\bf 476} (1996) 329
  [hep-th/9604002].

\bibitem{Wess:1971yu}
  J.~Wess and B.~Zumino,
  Phys.\ Lett.\ B {\bf 37} (1971) 95.

\bibitem{Luscher:1988sd}
  M.~L\"uscher,
  ``Selected topics in lattice field theory,''
  Conf.\ Proc.\ C {\bf 880628} (1988) 451.

\bibitem{Luscher:1998kn}
  M.~L\"uscher,
  Nucl.\ Phys.\ B {\bf 538} (1999) 515
  [hep-lat/9808021].

\bibitem{Niedermayer:1998bi}
  F.~Niedermayer,
  Nucl.\ Phys.\ Proc.\ Suppl.\  {\bf 73} (1999) 105
  [hep-lat/9810026].

\bibitem{Kikukawa:1998pd}
  Y.~Kikukawa and A.~Yamada,
  Phys.\ Lett.\ B {\bf 448} (1999) 265
  [hep-lat/9806013].

\bibitem{Fujikawa:1998if}
  K.~Fujikawa,
  Nucl.\ Phys.\ B {\bf 546} (1999) 480
  [hep-th/9811235].

\bibitem{Adams:1998eg}
  D.~H.~Adams,
  Annals Phys.\  {\bf 296} (2002) 131
  [hep-lat/9812003].

\bibitem{Suzuki:1998yz}
  H.~Suzuki,
  Prog.\ Theor.\ Phys.\  {\bf 102} (1999) 141
  [hep-th/9812019].

\bibitem{Chiu:1998qv}
  T.~-W.~Chiu and T.~-H.~Hsieh,
  hep-lat/9901011.

\bibitem{Abbott:1977in}
  L.~F.~Abbott, M.~T.~Grisaru and H.~J.~Schnitzer,
  Phys.\ Rev.\ D {\bf 16} (1977) 2995.

\bibitem{Curtright:1977cg}
  T.~Curtright,
  Phys.\ Lett.\ B {\bf 71} (1977) 185.

\bibitem{Inagaki:1978iu}
  H.~Inagaki,
  Phys.\ Lett.\ B {\bf 77} (1978) 56.

\bibitem{Majumdar:1980ej} 
  P.~Majumdar, E.~C.~Poggio and H.~J.~Schnitzer,
  Phys.\ Lett.\ B {\bf 93}, 321 (1980).

\bibitem{Nicolai:1980km} 
  H.~Nicolai and P.~K.~Townsend,
  Phys.\ Lett.\ B {\bf 93}, 111 (1980).

\bibitem{Hagiwara:1979pu}
  T.~Hagiwara, S.~-Y.~Pi and H.~-S.~Tsao,
  Annals Phys.\  {\bf 130} (1980) 282.

\bibitem{Hagiwara:1980ys} 
  T.~Hagiwara, S.~-Y.~Pi and H.~S.~Tsao,
  Phys.\ Lett.\ B {\bf 94}, 166 (1980).

\bibitem{Kumar:1982ng} 
  S.~Kumar and Y.~Fujii,
  Prog.\ Theor.\ Phys.\  {\bf 68}, 294 (1982).

\bibitem{Nakayama:1983qt} 
  R.~Nakayama and Y.~Okada,
  Phys.\ Lett.\ B {\bf 134}, 241 (1984).

\bibitem{Howe:1990pz}
  P.~S.~Howe, U.~Lindstr\"om and P.~White,
  Phys.\ Lett.\ B {\bf 246} (1990) 430.

\bibitem{Blasi:1995vt}
  A.~Blasi and N.~Maggiore,
  Mod.\ Phys.\ Lett.\ A {\bf 11} (1996) 1665
  [hep-th/9511068].

\bibitem{Weinberg:1996kr}
  S.~Weinberg,
  ``The quantum theory of fields. Vol. 2: Modern applications,''
  Cambridge, UK: Univ. Pr. (1996) 489 p

\bibitem{Barnich:2000zw}
  G.~Barnich, F.~Brandt and M.~Henneaux,
  Phys.\ Rept.\  {\bf 338} (2000) 439
  [hep-th/0002245].

\bibitem{White:1992wu}
  P.~L.~White,
  Class.\ Quant.\ Grav.\  {\bf 9} (1992) 413.

\bibitem{Maggiore:1994xw}
  N.~Maggiore,
  Int.\ J.\ Mod.\ Phys.\ A {\bf 10} (1995) 3937
  [hep-th/9412092].

\bibitem{Maggiore:1994dw}
  N.~Maggiore,
  Int.\ J.\ Mod.\ Phys.\ A {\bf 10} (1995) 3781
  [hep-th/9501057].

\bibitem{Catterall:2011pd}
  S.~Catterall, E.~Dzienkowski, J.~Giedt, A.~Joseph and R.~Wells,
  JHEP {\bf 1104} (2011) 074
  [arXiv:1102.1725 [hep-th]].

\bibitem{grassmann.m}
\url{http://people.brandeis.edu/~headrick/Mathematica/grassmann.m}

\end{thebibliography}



\end{document}